\documentclass[
twocolumn
,notitlepage
,floatfix
,balancelastpage
,aps
,pra
]{revtex4-1}

\usepackage{amsthm,amsmath,amssymb,amsfonts}
\usepackage{mathtools}
\usepackage{latexsym}
\usepackage{graphicx}  
\usepackage{bm}        
\usepackage{verbatim}
\usepackage{soul}
\usepackage{mathpazo}
\usepackage{enumerate}
\usepackage{comment}
\usepackage[caption=false]{subfig}
\usepackage[usenames,dvipsnames]{xcolor}
\usepackage{tikz}
\usepackage[framemethod=tikz]{mdframed}
\usepackage[pdftex,letterpaper=true,pagebackref=false]{hyperref}
\usepackage[nameinlink,capitalize]{cleveref}
\theoremstyle{plain}

\newtheorem{lemma}{Lemma}

\newtheorem{proposition}{Proposition}
\theoremstyle{definition}
\newtheorem{assumption}{Assumption}

\newcommand{\bra}[1]{\mbox{$\left\langle #1\right|$}}
\newcommand{\ket}[1]{\mbox{$\left|#1\right\rangle$}}

\newcommand{\ketbra}[2]{\mbox{$\left|#1\right\rangle\!\!\left\langle #2\right|$}}

\newcommand{\dket}[1]{\mbox{$\left|\left.#1\right\rangle\!\right\rangle$}}
\newcommand{\dbra}[1]{\mbox{$\left\langle\!\left\langle #1\right.\right|$}}

\newcommand{\dketdbra}[2]{\mbox{$|#1\rangle\!\rangle\!\langle\!\langle #2|$}}

\DeclareMathOperator{\Tr}{Tr}

\DeclareMathAlphabet{\bb}{U}{bbold}{m}{n}   
\newcommand{\id}{{\bb 1}}

\def\1#1{{\bf #1}}
\def\2#1{{\cal #1}}
\def\3#1{{\sl #1}}
\def\4#1{{\tt #1}}
\def\5#1{{\sf #1}}
\def\6#1{{\mathfrak #1}}
\def\7#1{{\mathbb #1}}

\def\bshade{\begin{mdframed}[hidealllines=true,backgroundcolor=blue!15]}
\def\eshade{\end{mdframed}}

\hypersetup{
pdftitle={Quantification and Characterization of Leakage Errors},
pdfauthor={Christopher J. Wood},
	pdfnewwindow=true, 
	colorlinks=true, 
	linkcolor=MidnightBlue,
	citecolor=Magenta,
	urlcolor=RoyalBlue
}

\begin{document}

\title{Quantification and Characterization of Leakage Errors}

\date{April 10, 2017}
\author{Christopher J. Wood}
\email{cjwood@us.ibm.com}
\affiliation{IBM T.J. Watson Research Center, Yorktown Heights, NY 10598, USA} 

\author{Jay M. Gambetta}
\affiliation{IBM T.J. Watson Research Center, Yorktown Heights, NY 10598, USA}

\begin{abstract}

We present a general framework for the quantification and characterization of leakage errors that result when a quantum system is encoded in the subspace of a larger system. To do this we introduce new metrics for quantifying the coherent and incoherent properties of the resulting errors, and we illustrate this framework with several examples relevant to superconducting qubits. In particular, we propose two quantities: the leakage and seepage rates, which together with average gate fidelity allow for characterizing the average performance of quantum gates in the presence of leakage and show how the randomized benchmarking protocol can be modified to enable the robust estimation of all three quantities for a Clifford gate set. 
\end{abstract}

\maketitle

\section{Introduction}
\label{sec:intro}

Accurate characterization of errors is critical for verifying the performance of quantum devices, and for prioritizing methods of error correction to improve the performance of quantum devices. In recent years there has been great progress in improving the performance of many types of qubits, with average gate fidelities exceeding $\sim\!\!\!0.999$ for 1-qubit and $\sim\!\!\!0.99$ for 2-qubit gates being reported in superconducting qubits~\cite{Barends2014nat,Sheldon2016pra2} and trapped ions~\cite{Benhelm2008nat,Ballance2016prl}. 
As thermal relaxation coherence times ($T_1$) increase, it becomes critical to quantify and measure the leading errors limiting further improvements in gate fidelities. In many quantum systems so-called leakage errors are an important factor for further gate optimization. These types of errors result from the state of a quantum system \emph{leaking out} of a pre-defined subspace to occupy an unwanted energy level. These types of errors are of particular importance in the context of fault-tolerant quantum error correction as they require significant additional resources to correct in a fault-tolerance manner over standard errors, and can greatly impact the fault tolerance-threshold of certain codes~\cite{Aliferis2007qic,Fowler2013pra,Suchara2015qic}. Furthermore even when very-weak or short-lived as the system returns to the desired subspace,  these types of interactions can result in significant logical errors, such as the AC-stark shift observed in the control of superconducting qubits \cite{Motzoi2009prl,Lucero2010pra}.

Leakage errors may be present in any quantum system where a qubit is encoded in a subspace of a larger quantum system as is the case for many qubit architectures including superconducting qubits~\cite{Gambetta2017nqi}, quantum dots~\cite{Divincenzo2000nat}, and trapped-ions~\cite{Haffner2008pr}. Though there has been significant interest in the characterization and suppression of leakage in quantum systems, current methods are highly architecture dependent~\cite{Lucero2010pra,Chow2010pra,Motzoi2009prl,Gambetta2011pra,Medford2013nat,Wardrop2014prb,Mehl2015prb2}, and while progress has been made (Eg. see \cite{Wallman2016njp,Chen2016prl}), there is not yet a general framework for quantifying and characterizing the relevant parameters of interest for an experimenter.

In this paper our goal is three-fold: in \cref{sec:leakage-errors,sec:coherent-leakage} we develop a unified framework for quantifying leakage errors which may occur in quantum systems; in \cref{sec:leakage-rb} we extend randomized benchmarking (RB)~\cite{Knill2008pra,Magesan2011prl} to a \emph{leakage randomized benchmarking} (LRB) protocol that estimates leakage errors in addition to average gate fidelities; in \cref{sec:leakage-models} we explore several canonical examples of leakage errors which may occur in quantum systems.
To quantify leakage, in \cref{sec:leakage-errors} we introduce measures for \emph{leakage in quantum states}, and for the quantum gate we introduce two measures which we call the \emph{leakage rate} $(L_1)$, and \emph{seepage rate} $(L_2)$. The leakage rate quantifies the average population lost from a quantum system of interest to states outside the computational subspace, while the seepage rate quantifies the return of population to the system from those states. To experimentally estimate these quantities we describe the leakage randomized benchmarking protocol (LRB) in \cref{sec:leakage-rb} and illustrate it's application with a numerical simulation of a transmon superconducting qubit. We note that this protocol and a similar approach have recently been demonstrated experimentally in Ref. \cite{McKay2016arx} and Ref. \cite{Chen2016prl} respectively.
In \cref{sec:coherent-leakage} we discuss the special case of \emph{coherent leakage errors} and introduce measures for the \emph{coherence of leakage} in quantum states and channels. While these measures cannot be directly estimated using the LRB protocol we prove bounds on these quantities in terms leakage and seepage. We note that in previous work the combined leakage-seepage rates $L_1+L_2$ was referred to as a measure of coherence of leakage~\cite{Wallman2016njp}, however this is a misnomer as leakage and seepage can result from purely incoherent thermal relaxation processes. We demonstrate and discuss this in \cref{sec:leakage-models} along with several other examples of leakage models including logical leakage errors, unitary leakage, thermal leakage, and multi-qubit leakage.

\subsection*{Comparison to previous work}

There have been previous proposals for generalizing RB to account for leakage~\cite{Epstein2014pra,Chasseur2015pra,Chen2016prl}, and also for related benchmarking protocols to explicitly quantify leakage instead of average gate fidelity~\cite{Wallman2016njp}. The  difference between our work and previous protocols is that it is designed to address the following experimental considerations
\begin{enumerate}
\item Allows for the robust estimation of the leakage rate $L_1$, seepage rate $L_2$, and average gate error $E$ of a Clifford gate set.
\item Can be implemented in any system capable of implementing RB with only minor modification.
\item The fitting model for parameter estimation from RB data is a single exponential decay model.
\end{enumerate}
The LRB protocol is essentially equivalent to the method recently used \cite{Chen2016prl,McKay2016arx} for characterization of leakage in a superconducting qubit. However, while that work relied on the assumption of a phenomenological decay model and direct measurement of leakage levels, we provide a more rigorous derivation of the decay model and discuss the assumptions for its validity. Our method also can be implemented without direct measurement of the leakage levels. 

To contrast our approach  with other previous work, in \cite{Epstein2014pra,Chasseur2015pra} they consider the decay model derived from RB in the presence of leakage for the case of a qubit or multi-qubit system respectively and hence this satisfies condition 2. This fails condition 1 as it does not provide a means for estimating the $L_1$ or $L_2$, only the gate error. Further, since the resulting decay model is a bi-exponential in the single qubit case, and a multiple-exponential in the multi-qubit case it fails to satisfy condition 3. The proposals of robust estimation of leakage in \cite{Wallman2016njp} satisfies conditions 2 and 3, and in particular the decay model used is equivalent to one part of our proposed protocol. However, it fail condition 1 as the characterization parameter reported by this protocol is equivalent to the sum of what we define as the leakage and seepage rates. It is critical to estimate these two quantities separately for the characterization of a quantum gate set in the presence of leakage, which we demonstrate with an example.

\section{Quantifying Leakage Errors}
\label{sec:leakage-errors}

Leakage in a quantum system can be modeled by treating the system of interest as a subspace of a larger quantum system in which the full dynamics occur. We will call the $d_1$-dimensional subspace of energy levels in which ideal dynamics occur the \emph{computational subspace}, labeled by $\2X_1$. The $d_2$ dimensional subspace of all additional levels that the system may occupy due to leakage dynamics will be called the \emph{leakage subspace}, labeled by $\2X_2$. Thus the full state space of the system is described by a $d_1+d_2$ dimensional direct-sum state space $\2X = \2X_1\oplus \2X_2$. We define the \emph{state leakage} ($L$) of a density matrix $\rho \in D(\2X)$ by
\begin{equation}
L(\rho) = \Tr[\id_2 \rho] = 1-\Tr[\id_1\rho],
\label{eq:state-leakage}
\end{equation}
where $\id_1$ and $\id_2$ denote the projectors onto the subspaces $\2X_1$ and $\2X_2$ respectively. 

For a quantum state to exhibit leakage it must be introduced into the system by some physical process, for example thermal-relaxation or imperfect control errors. In general we refer to any system dynamics which result in a change of the state leakage of quantum system as a \emph{leakage error}.  In the quantum circuit paradigm imperfect quantum operations may be described mathematically by completely-positive trace preserving maps (CPTP), and thus a leakage error is a special class of CPTP map that couples the computational and leakage subspaces. Let $\2E$ be a CPTP map describing a leakage error. We can quantify the leakage error in $\2E$ by how it changes the state leakage of an input state. However, unlike with state leakage we will require two metrics to distinguish between leakage errors which transfer population \emph{to}, and population \emph{from}, the leakage subspace. We call these errors \emph{gate leakage} and \emph{gate seepage} respectively.

Much like with average gate fidelity to quantify typical gate errors we will generally be interested in the average and the worst case gate leakage and seepage. Thus we define the \emph{leakage rate} $L_1$, and \emph{seepage rate} $L_2$ of a channel $\2E$ to be the average channel leakage and channel seepage respectively:
\begin{align}
L_{1}(\2E) 
	&=  	\int \!
	d\psi_1 L\Big(\2E(\ketbra{\psi_1}{\psi_1})\Big)  =L\left(\2E\left(\frac{\id_1}{d_1}\right)\right)
	\label{eq:leakage-rate}
	\\
L_{2}(\2E) 
	&=  1-\int \!
	d\psi_2 L\Big(\2E(\ketbra{\psi_2}{\psi_2})\Big) = 1- L\left(\2E\left(\frac{\id_2}{d_2}\right)\right)
	\nonumber
\end{align}
where the averages are over all state in the computational subspace $(\ket{\psi_1})$, and leakage subspace $(\ket{\psi_2})$ respectively.

The worst case gate leakage and seepage rates which require maximizing, rather than averaging, over all input states. We note however that we may bound the worst-case quantities by average rates as was shown in \cite{Wallman2016njp}:
\begin{align}
d_1 L_1(\2E)  \ge&  L\big(\2E(\rho_1)\big) 	\label{eq:worst-p1}\\
 d_2 L_2(\2E) \ge& 1-L\big(\2E(\rho_2)\big) \label{eq:worst-p2}
\end{align}
where $d_j$ is the dimension of $\2X_j$, and and $\rho_1(\rho_2)$ is an arbitrary state in, the computational (leakage) subspace.

In addition to characterizing the amount of leakage introduced by an imperfect gate, it is also necessary to characterize the performance of the gate within the computational subspace. A commonly used measure of gate error is the average gate infidelity $E=1-\overline{F}$ where $\overline{F}$ is the average gate fidelity
\begin{equation}
\overline{F}(\2E) = \int d\psi\bra{\psi} \2E(\ketbra{\psi}{\psi})\ket{\psi}.
\end{equation}
In the presence of leakage we define $\overline{F}$ by only averaging over \emph{states within the computational subspace}:
\begin{align}
\overline{F}(\2E)
	&= \int d\psi_1 
	\bra{\psi_1} \2E(\ketbra{\psi_1}{\psi_1})\ket{\psi_1} \label{eq:fid-def}\\
	&= \frac{d_1F_{\text{\scriptsize{pro}}}(\2E)+1-L_1}{d_1+1} \label{eq:leakage-fid}
\end{align}
where we have expressed $\overline{F}$ in terms of the \emph{process fidelity} of $\2E$ with the identity channel on the computational subspace:
\begin{align}
F_{\text{pro}}(\2E) = \frac{1}{d_1^2} \Tr[(\id_1\otimes\id_1)\2S_{\2E}]
\end{align}
where $\2S_{\2E}$ is the superoperator representation of the map $\2E$~\cite{Wood2015qic}. 

We suggest that the goal of a partial characterization protocol of leakage errors is to extract the three parameters $L_1,L_2$ and $E$. This is a major difference between our approach and the protocol in \cite{Wallman2016njp} which only aims to learn a single parameter equivalent to the joint leakage-seepage rate $L_1+L_2$. Knowledge of the combined leakage rate is insufficient to accurately quantify the gate error, and in addition the relationship between $L_1$ and $L_2$ can vary greatly depending on the noise process causing leakage.  Three specific cases are: \emph{erasure leakage errors} where $L_2=0$ and hence any leaked population is irretrievably lost; \emph{thermal relaxation leakage errors} where $L_2 \gg L_1$ if the computational subspace is the low energy subspace of the system; \emph{unital leakage errors} (of which unitary leakage errors are a subset) where the leakage and seepage rates are no longer independent and satisfy the equation $d_1 L_1 = d_2 L_2$~\footnote{This result follows directly from the definitions for leakage and seepage. For a given CPTP map $\2E$ we have $d_1 L_1 (\2E)= \Tr[\id_2 \2E(\id_1)] = \Tr[\id_2 \2E(\id)] - d_2(1-L_2(\2E))$. Now, if $\2E$ is unital then $\Tr[\id_2 \2E(\id)]=\Tr[\id_2] =d_2$, and hence
$d_1 L_1 (\2E)=d_2 L_2 (\2E)$.}. 
We elaborate on this when giving several example noise models in \cref{sec:leakage-models}. We also note that the definitions of leakage and seepage can be generalized for leakage to multiple different leakage subspaces by further partitioning the leakage subspace into several different subspaces. We discuss this in more detail in \cref{sec:multi-leakage}.

\section{Characterizing Leakage Errors}
\label{sec:leakage-rb}

We now show how the randomized benchmarking (RB) protocol can be generalized to include estimation of the leakage and seepage rates in addition to the average gate fidelity for a Clifford gate set. We call this generalized protocol \emph{leakage randomized benchmarking (LRB)}.

The basic requirements for LRB are the implementation of a set of gates which form a 2-design on the computational subspace, the typical set being the Clifford gates, and the ability to perform a set of measurements $\{M_0,\hdots M_{d_1-1}\}$ which may be used to estimate the populations of a set of basis states in $\2X_1$. By summing over all measurements we may implement the projection $\sum_j M_j = \id_1$, and so this allows for the estimation of the population in $\2X_1$, and hence of the state leakage $L$. In the following we will assume that this is done with respect to the computational basis. In this sense it is equivalent to RB, but with additional measurements and different post-processing of the acquired data. With this requirement the LRB protocol is as follows:
\begin{enumerate}
\item Choose a random sequence of $m$ Clifford gates $i_m = \2C_m \circ \hdots \circ \2C_1$ and compute the RB recover operator corresponding to $\2C_{m+1} = \2C_1^\dagger \circ\hdots \circ \2C_m^\dagger$ to obtain the RB sequence $i_m^\prime = \2C_{m+1}\circ i_m$.
\item Prepare the system in an initial state $\rho_0=\ketbra00 \in D(\2X_1)$, apply the sequence $i_m^\prime$, and perform a measurement to obtain an estimate of the probabilities $p_j(i_m^\prime) = \Tr[M_j i_m^\prime(\rho_0)]$, where $j=0,\hdots, d_1-1$.
\item Sum the probabilities $p_j(i_m^\prime)$ to obtain an estimate of the population in $\2X_1$: $p_{\id_1}(i_m^\prime) = \sum_j p_j(i_m^\prime) = \Tr[\id_1 i_m^\prime(\rho_0)]$. 
\item Repeat Steps 1-3 $K$ times to obtain an estimate of the average over all Clifford sequences $i_m$: 
	$p_j(m)= \1E_{i_m^\prime}\left[ p_j(i_m^\prime)\right]$, 
	$p_{\id_1}(m)= \1E_{i_m^\prime}\left[ p_{\id_1}(i_m^\prime)\right]$.
\item Repeat 1-4 for different sequence lengths $m$.
\item Fit 
	$p_{\id_1}(m)$ to the decay model
	\begin{equation}
	p_{\id_1}(m) = A + B \lambda_1^m
	\label{eq:lrb-leak}
	\end{equation}
	with $0\le A,B$ to obtain estimated values for $A, B$ and $\lambda_1$. Compute the estimate of the average leakage and seepage rates of the gate set as
	\begin{align}
	L_1(\2E) &= (1-A)(1-\lambda_1)	\label{eq:fit-p1}\\
	L_2(\2E) &= A(1-\lambda_1).	\label{eq:fit-p2}
	\end{align}
	Note that in practice one may put tighter bounds on the expressions based on estimates for leakage rates using 
	\begin{align}
	A &\approx \frac{L_2}{L_1+ L_2} \\
	B &\approx \frac{L_1}{L_1+L_2} + \epsilon_{\text{spam}}\\
	\lambda_1 &= 1 - L_1 - L_2.
	\end{align}
\item Using the fitted value of $\lambda_1$ fit $p_0(m)$ to the decay model
	\begin{align}
	p_0(m) &= A_0 + B_0 \lambda_1^m + C_0 \lambda_2^m
	\label{eq:lrb-fid}
	\end{align}
	where $0\le A_0 \le A, 0\le C_0 \le1, 0\le A_0 +B_0+C_0 \le 1$, to obtain an estimate of the average gate fidelity of the gate set  by
	\begin{align}
	\overline{F} &= \frac{1}{d_1}\Big[(d_1-1)\lambda_2
			+ 1-L_1\big].  \label{eq:lrb-fid}
	\end{align}	
If leakage is weak ($\lambda_2 \ll \lambda_1, B\ll A$ ), a more robust fit  can be obtained from fitting directly to a RB model $p_0(m) = 	A_0 +C_0 \lambda_2^m$.
\end{enumerate}
See \cref{sec:lrb-decay-model} for the derivation of the decay models \cref{eq:lrb-leak,eq:lrb-fid} in Steps 7 and 8.

The LRB protocol subsumes the various decay models previously presented in the literature. The modified RB decay model presented in \cite{Epstein2014pra} equivalent to the model in \cref{eq:lrb-fid} for the case of leakage to a single level. The protocol presented in \cite{Wallman2016njp} is based of a 1-design average rather than a 2-design and is equivalent to our model in the case where the recovery operation is not included hence replacing the sequence $i_m^\prime$ with $i_m$ in Steps 2-5 in the LRB protocol. In this case the resulting decay model is equivalent to the one given in \cref{eq:lrb-leak}. Finally the phenomenological decay model assumed in \cite{Chen2016prl} is equivalent to an implementation of the LRB protocol with direct measurements of the leakage subspace and thus our work provides a theoretical justification of the assumptions and validity of this model. 

In the case where one has the ability to directly measure populations of the leakage subspace LRB Steps 1,2,4,5,7 are implemented the as is, but step 3 is replaced with a measurement set chosen to correspond to estimates of the population of the leakage subspace with respect to some basis
\[
\{N_j = \ketbra{j}{j} : j = 1+d_1,..., d_2 + d_1-1\}.
\]
Following this Steps 6 is replaced with fitting to the model
\begin{align}
p_{\id_2}(m)
	&= 1-p_{\id_1}(m)= 1-A-B\, \lambda_1^m \label{eq:lrb-leak-direct}
\end{align}
This is of course most beneficial when the dimension of the leakage subspace is less than the computational subspace, and in particular when the leakage subspace is 1-dimensional gives a method of quickly estimating the leakage rates. 

\subsection{Simulation of LRB for a Superconducting Qubit}
\label{sec:lrb-sim}

We now demonstrate the application of the LRB protocol with a simulation of the a superconducting qubit, and note that this protocol has also implemented experimentally in Ref.~\cite{McKay2016arx}. A superconducting qubit is a weakly anharmonic oscillator which to a good approximation can be described by truncating the system to a three dimensional Hilbert space. In this case the qubit computational subspace is spanned by the states $\{\ket0,\ket1\}$, and the leakage subspace is a single level $\ket2$. The Hamiltonian for the system in the resonance frame of the $E_1-E_0$ energy separation is given by
\begin{align}
H(t) =& H_0 + H_c(t) \\
H_0 =& -\delta \ketbra{2}{2} \\
H_c(t) =& \frac12\Omega_x(t) H_x + \frac12\Omega_y(t) H_y \\
H_x =& \ketbra01 + \sqrt2\ketbra12 + h.c.\\
H_y =& -i\ketbra01 -i\sqrt{2}\ketbra12 + h.c.
\end{align}
where $\delta$ is the anharmonicity, and $H_c(t)$ is the time-dependent control Hamiltonian. We will consider a family of $y$-only DRAG corrected pulse shapes~\cite{Gambetta2011pra} where the $x$-drive component $\Omega_x(t)$ is a truncated Gaussian pulse, and the $y$-drive component is given by the scaled derivative
\begin{align}
\Omega_y(t) & = -\frac{\alpha}{\delta} \frac{d}{dt} \Omega_x(t). \label{eq:drag}
\end{align}
To include the effects of thermal relaxation, we model dissipation of the system as cavity relaxation to an equilibrium state with average photon number $\overline{n}$. This is described by the Markovian photon-loss dissipator~\cite{Wiseman2009book}\begin{align}
\2D_c &= \kappa\left(1+\overline{n}\right) D[a] + \kappa\,\overline{n}\, D[a^\dagger]  \label{eq:diss}\\
D[a]\rho &= a\rho a^\dagger -\frac12\{a^\dagger a,\rho\},
\end{align}
where $\kappa$ is the relaxation rate of the system, $\overline{n}$ is the average thermal photon number, and $a,a^\dagger$ are the truncated creation and annihilation operators. To compute the superoperator for a given control pulse we solve the Lindblad master equation
\begin{equation}
\frac{d\rho(t)}{dt} = - i [H(t),\rho] + \2D_c \rho.
\end{equation}
over the time-dependent control pulse.

\begin{figure}[t]
\centering
\includegraphics[width=0.95\columnwidth]{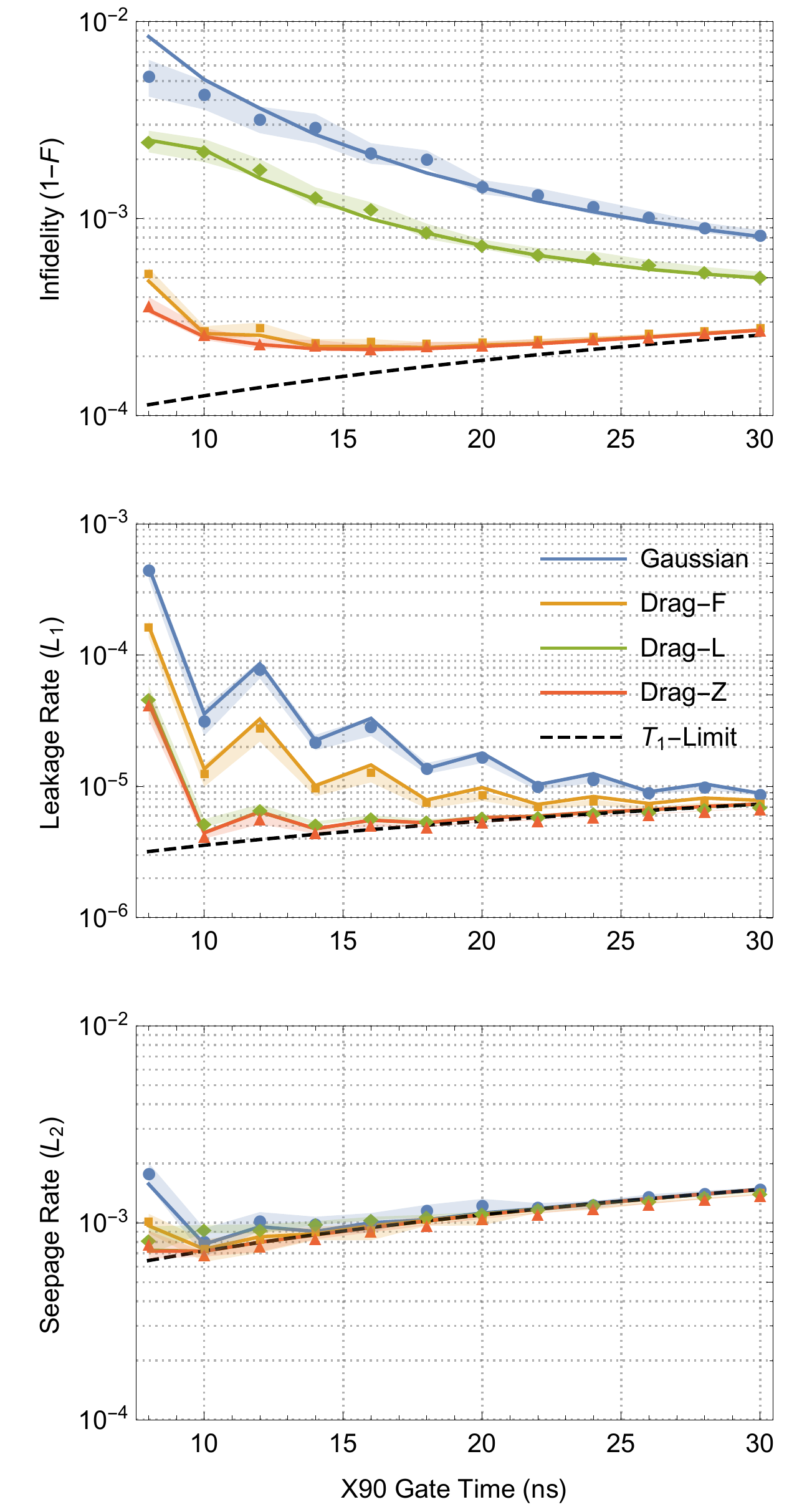}
\caption{Average gate infidelity (top), leakage rate (middle) and seepage rate (bottom) averaged over the single qubit Clifford gates versus gate time of the component pulses. Data points are the fitted estimates from simulation of the LRB protocol with 100 seeds for each length-$m$ sequence, with shaded regions representing the 95\% confidence interval of the fit. The solid lines are theoretical values computed directly from the superoperators used for simulation, with the theoretical $T_1$-limit given by pure thermal relaxation noise only.}
\label{fig:lrb-leakage}
\end{figure}

\begin{figure}[t]
\centering
\includegraphics[width=\columnwidth]{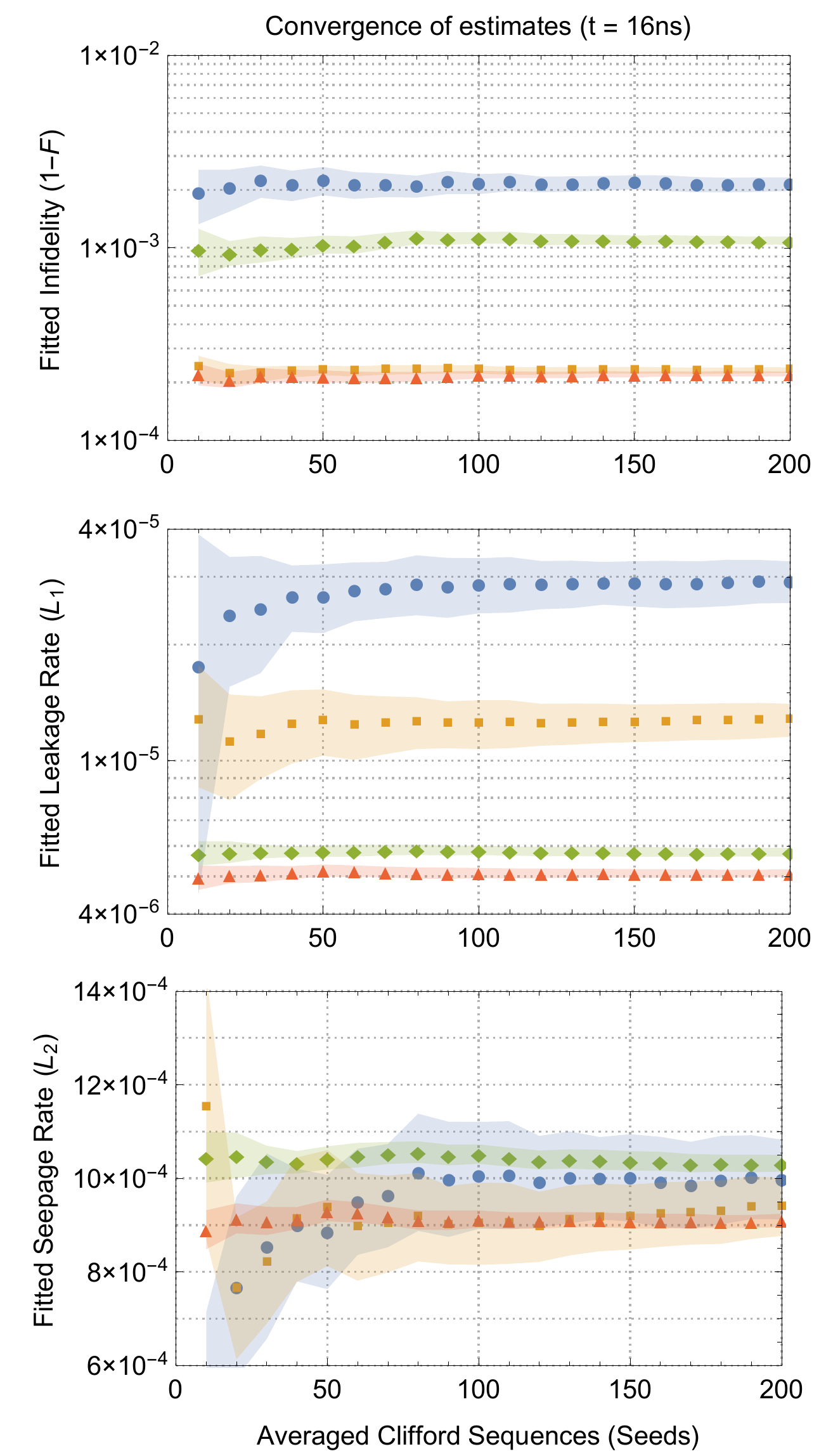}
\caption{Convergence of estimate of average gate infidelity (top), leakage rate (middle) and seepage rate (bottom) vs number of random Clifford sequences used to obtain average. Data points are the fitted estimates from simulation of the 16ns component pulse from \cref{fig:lrb-leakage}, with shaded regions representing the 95\% confidence interval of the fit.}
\label{fig:lrb-seeds}
\end{figure}

\begin{figure*}[t]
\centering
\includegraphics[width=\textwidth]{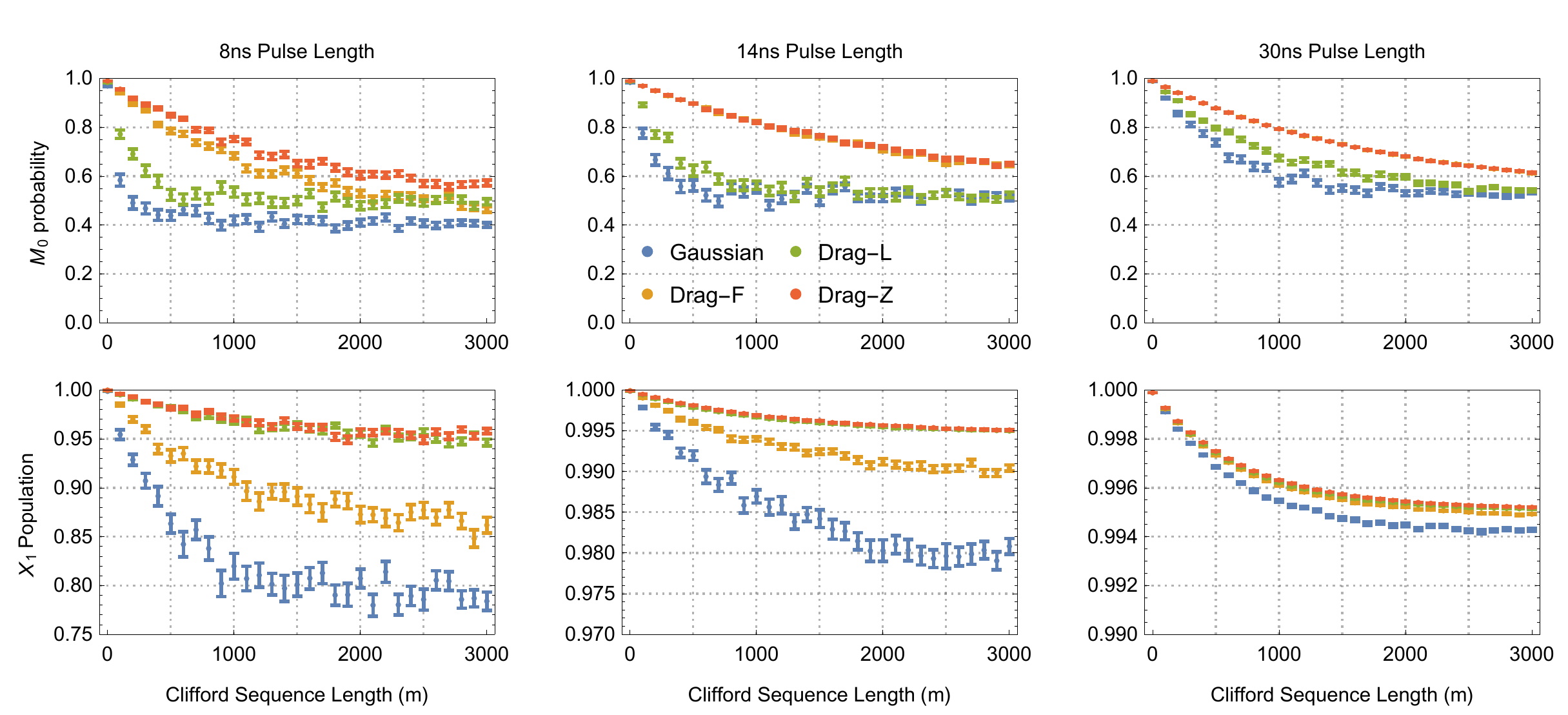}
\caption{Example decay data for the simulated LRB experiment in \cref{fig:lrb-leakage}. The top row shows the $p_0$ decay curves of the $\ket{0}$ state measurement typically used in RB, while the bottom row shows the $p_{\id_1}$ decay curve for the trace of the qubit computational subspace. The left, middle, and right columns are data for 8, 14, and 30ns length component pulses for generating the Clifford gates respectively.}
\label{fig:lrb-data}
\end{figure*}

For our simulation we compare the LRB estimates of mean leakage rate, and seepage rate, and average gate infidelity for a single-qubit Clifford gate set to the theoretical values computed directly from the Clifford gate superoperators. The noisy Clifford gate set was generated by simulating calibrated $\pm \pi/2$ $X$ and $Y$ rotation pulses for a transmon qubit with anharmonicity $\delta/2\pi = -300$~MHz, and thermal relaxation modeled as cavity dissipation with relaxation time constant 
$\kappa = 10$~ kHz, and an average photon number at thermal equilibrium of $\overline{n}=0.01$. To simulate leakage that occurs during measurement of the qubit we allow the system to evolve under the dissipator in \cref{eq:diss} for a typical measurement acquisition time of $5\mu s$ following the final pulse. The $X$,$Y$ pulses were simulated for a truncated gaussian pulses of lengths ranging from 8ns to 30ns with a 4ns spacing before and after each pulse. We compare four different pulse types: \emph{GAUSS} which has DRAG parameter $\alpha=0$, \emph{DRAG-F} where $\alpha$ is optimized to maximize average gate fidelity ($\alpha \approx 0.5$), \emph{DRAG-L} where has $\alpha$ is optimized to minimize the leakage rate ($\alpha\approx 1$), and \emph{DRAG-Z} which uses the leakage optimized $\alpha$ from DRAG-L but also uses and optimized Z-frame changes to maximize average gate fidelity as introduced in Ref.~\cite{McKay2016arx}. The LRB protocol was simulated using Clifford sequence lengths of $m=1, 101,201,\hdots, 3001$ averaging over 100 random seeds for each $m$.  The results are shown in \cref{fig:lrb-leakage}. To illustrate the error in the estimates \cref{fig:lrb-seeds} shows the fitted values and 95\% confidence intervals of the fits versus the number of random Clifford sequences (seeds) averaged over for each length $m$ for the case of 14ns pulse used in \cref{fig:lrb-leakage}. Examples of simulated LRB  data used for fitting in \cref{fig:lrb-leakage} are shown in \cref{fig:lrb-data}.

The results in \cref{fig:lrb-leakage} show good agreement with the LRB fitted estimate and the directly computed theoretical values for leakage, seepage and infidelity. We note that due to the leakage rate being over an order of magnitude smaller than the average gate fidelity we are able to fit the fidelity RB curve to a single exponential as with standard RB.
We see that the DRAG-Z and DRAG-F have the greatest improvement of gate infidelity, approaching the $T_1$ limit for longer pulses, with DRAG-Z only having an advantage for pulse times shorter than 14ns. The DRAG-L pulse shows a marginal improvement over the GAUSS pulse. This is because the dominant error is a leakage induced phase error, rather than the leakage rate itself, and hence is not corrected by the DRAG-L calibration. When comparing leakage rates, we see that for all pulses the leakage rates are typically 1 to 2 orders of magnitude below the infidelity. Both the DRAG-L and DRAG-Z pulses saturate the $T_1$ limit on leakage rates $L_1$ for pulses longer than 14ns, while the DRAG-F and GAUSS pulses only approach this limit for much longer pulses. For the case of seepage, as was discussed in \cref{sec:cavity-leakage}, we find that it is completely dominated by the thermal seepage due to $T_1$ relaxation.

These simulations show that in the case of single-qubit gates in superconducting qubits, leakage induced errors are much more important for device optimization than the leakage rates themselves. Leakage and seepage rates are limited by $T_1$ relaxation methods, however if minimizing leakage errors significantly below the average gate fidelity is a requirement for a fault-tolerant codes, for example, then saturating this leakage $T_1$ limit in combination with the infidelity limit may be a desirable goal for control design. In this case simple half-DRAG pulses alone are not sufficient, and one must use other methods to optimize calibration to remove both the leakage induced phase error and suppress leakage rates such as the DRAG-Z pulse which uses DRAG in combination with Z-rotations to remove phase errors presented in Ref.~\cite{McKay2016arx}. Another method is to use DRAG in combination with a detuned drive term as done in Ref.~\cite{Chen2016prl}. Theoretical approaches to systematically design new control pulses to achieve this have also been recently proposed~\cite{Ribeiro2017prx}.

\subsection{Assumptions of the LRB Protocol}
\label{sec:lrb-assumptions}

The derivation of the LRB decay models in \cref{eq:lrb-leak,eq:lrb-fid} require two key assumptions in addition to the requirements of RB. The first assumption is that when twirling over Clifford gates on the computational subspace there is enough randomness induced on the leakage subspace to implement a unitary 1-design average, thus depolarizing the conditional state on the leakage subspace. In many system of interest, for example superconducting qubits, leakage is typically weak and restricted to a 1-dimensional subspace that is off-resonance with the computational subspace levels. In this case the depolarizing requirement is trivially satisfied, as depolarization is implemented by the random phases accrued due to off-resonant evolution. 

If the leakage subspace is not sufficiently depolarized then non-Markovian effects may appear in the case where the leakage dynamics are coherent due to coherences between the leakage and computational subspace. We discuss coherent leakage in greater detail in \cref{sec:coherent-leakage}, and explore an example of when this twirling assumption breaks down in \cref{sec:unitary-dlm}. The key feature of insufficient twirling of the leakage subspace is oscillations in the leakage decay model. We can see this, for example, in the LRB data for the 8ns pulse decay curves in \cref{fig:lrb-data}, which was the gate set with the largest leakage rate in our simulation. In the case where the leakage subspace is only partially depolarized.

\section{Coherent Leakage Errors}
\label{sec:coherent-leakage}

The leakage metrics in \cref{sec:leakage-errors} all measure incoherent properties of a quantum system. In particular the population of the leakage subspace used as the definition of state leakage does not inform us about coherences that may exist between states in the computational and leakage subspaces. Our restriction to these leakage metrics is a practical one as estimation of coherent properties of leakage is considerably more difficult. In the ideal case the LRB protocol acts to project our coherent leakage terms, and direct estimation requires the ability to directly measure coherences between the leakage and computational subspaces which. Nevertheless, we now present a theoretical framework for quantifying coherent leakage errors and show how these quantities may be bounded by the average leakage quantities from \cref{sec:leakage-errors}, and later in \cref{sec:unitary-leakage} we explore a simple example of a unitary coherent leakage error.

\subsection{Coherence of Leakage}
\label{sec:coherence-of-leakage}

Consider a leakage system with state space $\2X= \2X_1\oplus\2X_2$, with identity operators $\id_1$ and $\id_2$ which project a state on the computational and leakage subspaces respectively. We can define a subspace consisting of all states of the form $\rho = (1-p_l)\rho_1 + p_l \rho_2$ which we call the \emph{incoherent leakage subspace (ILS)} as it is an incoherent mixtures of states in the leakage and computational subspaces. The projector onto the ILS is given by the CPTP map 
\begin{align}
\2P_I &= \2I_1 + \2I_2 \\
\2P_I(\rho) &= \id_1\rho\id_1 + \id_2\rho\id_2
\label{eq:ils-proj}
\end{align}
where $\2I_1,\2I_2$ are the identity projection channels for the computational and leakage subspaces respectively. 

For a state $\rho$, the the orthogonal subspace to the ILS  is traceless and consists of \emph{only} the coherent superposition terms between states in the computational and leakage subspace. We will call this subspace the \emph{coherent leakage subspace (CLS)}. The projector onto the CLS is given by
\begin{align}
\2P_C &= \2I - \2P_I \\
\2P_C(\rho) &= \id_1\rho\id_2 + \id_2\rho\id_1
\label{eq:cls-proj}
\end{align}
where $\2I$ is the identity channel on the full Hilbert space. We may use the CLS projection to define a measure of the quantify the coherence of leakage in a density matrix. We define the \emph{coherence of leakage} of a state $\rho$ to be
\begin{equation}
CL(\rho) = \| \2P_C(\rho)\|_1 = \| \rho - \2P_I(\rho) \|_1.
\label{eq:leak-c}
\end{equation}
While we could use any suitable matrix norm, the choice of the 1-norm is to give an operational interpretation of $CL(\rho)$ via Helstrom's theorem. 
For example, consider a pure state $\rho= \ketbra{\psi}{\psi}$ consisting of a superposition of states in the leakage and computational subspaces. If the leakage of $\rho$ is given by 
\begin{equation}
L(\rho)=\bra\psi\id_2\ket\psi=p_l
\end{equation}
we may write 
\begin{equation}
\ket{\psi}=\sqrt{1-p_l}\ket{\psi_1}+\sqrt{p_l}\ket{\psi_2}
\end{equation}
where $\ket{\psi_j}\in\2X_j$. Hence we have
\begin{align}
\| \2P_C(\ketbra\psi\psi)\|_1
&= 2\sqrt{p_l(1-p_l)}. 
\label{eq:cl-bound}
\end{align}
The norm in \cref{eq:cl-bound} equals 1 when $p_l=1/2$. As one might expect this shows that our ability to distinguish coherent leakage from purely incoherent leakage is maximized when there is an equal superposition of states in $\2X_1$ and $\2X_2$. If $p_l=0$ or 1, so that the state is entirely in the computational or leakage subspace, then there can be no coherences between the leakage and computational subspaces and $\| P_C(\rho)\|_1=0$. 

While the trace distance of the CLS projection has a useful operational interpretation it cannot be directly measured from measurements on the computation subspace alone. We can, however, prove that \cref{eq:cl-bound} provides an upper bound on the coherence of leakage.
\begin{proposition}
\label{prop:coherence-bound}
Consider a density matrix $\rho\in \2L(\2X_1\oplus\2X_2)$. The coherence of leakage is upper bounded by
\begin{equation}
CL(\rho) \le  2\sqrt{p_l(1-p_l)} \nonumber
\label{eq:leak-c}
\end{equation}
where $p_l = L(\rho)$ is the leakage of $\rho$. 
\begin{proof} See \cref{proof:coherence-bound}.\end{proof}
\end{proposition}

As shown in \cref{eq:cl-bound} the bound in \cref{prop:coherence-bound} is saturated for a pure state $\rho$.

\subsection{Coherent Leakage Rates}
\label{sec:gate-leakage}

The definitions of leakage and seepage rates introduced in \cref{sec:leakage-errors} quantify the rates at which a CPTP error map $\2E$ increases or decreases the amount of state leakage of a given input state. We now consider how coherent leakage errors can be introduced into a system. Consider an arbitrary leakage error channel described by a CPTP map $\2E$. We may use the projectors for the ILS and CLS from \cref{eq:ils-proj,eq:cls-proj} to decompose $\2E$ into four channel components:
\begin{equation}
\2E = \2P_I\2E\2P_I  + \2P_I\2E \2P_C + \2P_C\2E\2P_I+ \2P_C\2E\2P_C.
\label{eq:chan-leak-decomp}
\end{equation}
The first term $\2E_I\equiv\2P_I\2E\2P_I$ is the trace preserving component of $\2E$, which we call the \emph{incoherent leakage component} of $\2E$. The remaining three terms all result in a traceless output operator. The incoherent leakage component may itself be expressed as $2\times 2$ block-mapping between the leakage and computational subspaces:
\begin{align}
\2E_I &= \2I_1 \2E \2I_1 + \2I_2 \2E \2I_1 + \2I_1 \2E \2I_2 +\2I_2 \2E \2I_2. \label{eq:incoh-decomp}
\end{align}
The TP property of $\2E$ allows us to write \cref{eq:incoh-decomp} in terms of the leakage and seepage rates $L_1,L_2$ as
 \begin{align}
\2E_I & 
	= (1-L_1)\2E_{11} + L_1 \2E_{21}+L_2\2E_{12} + (1-L_2) \2E_{22}
\label{eq:incoh-leak-comp}
\end{align}
where
\begin{equation}
\2E_{ij} = \frac{\2I_i \2E \2I_j}{\Tr[\id_i \2E(\id_j/d_j)]}.
\end{equation}
\cref{eq:incoh-leak-comp} shows that the incoherent leakage component $\2E_I$ is the only relevant term for estimating leakage and seepage rates as defined in \cref{sec:leakage-errors}, and under ideal situations the LRB protocol from \cref{sec:leakage-rb} projects an arbitrary error channel $\2E$ onto this component. If one doesn't project out the terms that allow for coherent leakage, then analogous to our definition of leakage and seepage rates, we can define two quantities to measure how much the coherence of leakage of a state is increased by leakage and seepage. We define the coherent leakage rate, and coherent seepage rate to be
\begin{align}
CL_1(\2E) 
	=& \int d\psi_1 CL\big(\2E(\ketbra{\psi_1}{\psi_2})\big)\\
CL_2(\2E) 
	=& \int d\psi_2 CL\big(\2E(\ketbra{\psi_1}{\psi_2})\big)\\
\end{align}
respectively, where $\ket{\psi_j}\in\2X_j$.
While this expression can be evaluated exactly for simple examples (such as the unitary leakage example in \cref{sec:unitary-leakage}), for more complicated examples we can always upper bound it by the leakage and seepage rates of the channel $\2E$ which we prove in \cref{prop:coh-leakage-bound}.

\begin{proposition}
\label{prop:coh-leakage-bound}
The coherent leakage rate $CL_1(\2E)$ and coherent seepage rate $CL_2(\2E)$ of a CPTP map $\2E$ are upper bounded by
\begin{equation}
CL_j(\2E) \le  2\sqrt{L_j(1-L_j)} \nonumber
\label{eq:leak-c}
\end{equation}
where $L_1, L_2$ are the leakage and seepage rates of $\2E$.
\begin{proof} See \cref{proof:coh-leakage-bound}.\end{proof}
\end{proposition}

We note that if one attempts to implement the LRB protocol, but does not successfully project onto the incoherent leakage channel component, the resulting decay model may exhibit oscillations due to the coherent leakage terms. We explore this with a simple example in \cref{sec:unitary-dlm}.

\section{Leakage Error Models}
\label{sec:leakage-models}

We now present some example models for leakage errors in quantum gates. These will cover simple logical models for erasure and depolarizing leakage for circuit simulations; dissipative leakage for modeling thermal leakage errors in physical qubits; and unitary leakage for leakage errors induced by quantum control.

\subsection{Logical Leakage Errors}
\label{sec:logical-leakage}

\subsubsection{Erasure Error}
\label{sec:erasure-error}

The simplest model for a leakage type error is an \emph{erasure error} where with some probability $p$ we completely \emph{lose} our qubit. This could, for example, correspond to an atom escaping a trap, or a photon escaping from a cavity or wave-guide. To model this in the leakage framework outlined in \cref{sec:leakage-errors} we can represent an erasure channel with erasure probability $p_l$ as a CPTP map
\begin{equation}
\2E(\rho) = (1-p_l)\rho + p_l \ketbra{\psi_2}{\psi_2}
\label{eq:loss}
\end{equation}
where $\ket{\psi_2} \in \2X_2$ is a state in the leakage subspace. In this case the leakage dynamics are entirely incoherent and we can think of the leakage subspace as a 1-dimensional system which keeps track of the lost population. The leakage and seepage rates for the erasure channel $\2E$ are given by $L_1=p_l$ and $L_2=0$.

If we measure the leakage after $m$ applications of the channel, then the leakage of the output state is given by
\begin{equation}
p_l(m) = L(\2E^{\circ m}(\rho)) = 1 - (1-p_l)^m
\end{equation}
This shows that the state leakage $p_l$ approaches 1 as $m$ increases and in the infinite limit we can say that its population is contained entirely in the leakage subspace.

\subsubsection{Depolarizing Leakage Extension}
\label{sec:depol-leakage}

Erasure errors are not a particularly realistic model for many architectures, such as spins, atomic systems, or superconducting qubits, where relaxation or other processes allow the higher energy levels to continue to interact with the computational subspace energy levels. We can consider erasure errors to be a subset of a more general leakage model which we call \emph{depolarizing leakage} in the limit where the seepage rate goes to 0.

Let $\2E_1$ be an arbitrary CPTP map on the computational subspace. We define the \emph{depolarizing leakage extension} (DLE) of $\2E_1$ to be the channel
\begin{equation}
\2E_L = (1-L_1) \2E_1 + L_1 \2D_{21} + L_2\2D_{12} + (1-L_2)\2D_{2} 
\label{def:leakage-ext}
\end{equation}
where $L_1, L_2$ are the leakage and seepage rates for the extension, $\2D_j \equiv \2D_{jj}$, and $\2D_{ij}$ is a completely depolarizing map between subspaces $\2L(\2X_i)$ and $\2L(\2X_j)$ given by
\begin{equation}
\2D_{ij}(\rho) = \Tr[\id_j \rho]\, \frac{\id_i}{d_i}, \quad i,j\in \{1,2\}.
\end{equation}

The DLE channel $\2E_L$ is a purely incoherent leakage error as it removes all information about the leakage dynamics \emph{except} for the leakage and seepage rates. The simplicity of this model could prove useful as a channel extension for including the effects of leakage in full characterization protocols such as gate set tomography~\cite{Merkel2013pra,Dehollain2016njp}. The leakage channel components $\2D_{12}$, $\2D_{21}$ act to remove any coherence of leakage in an initial state, and in combination with the completely depolarizing component $\2D_{2}$ on the leakage subspace ensures that there are no memory effects in the leakage subspace dynamics. This assumption ensures an exponential model for the state leakage under repeated applications of the DLE. 
\begin{lemma}\label{prop:dle-leak}
Let $\2E_L$ be a DLE and $\rho_1\in D(\2X_1)$ be a state in the computational subspace. Then the state leakage accumulation model for an initial state $\rho$ due to repeated actions of $\2E_L$ is given by
\begin{equation*}
L(\2E_L^m(\rho)) =  \frac{L_1}{L_1+L_2} -\left(\frac{L_1}{L_1+L_2}-p_l\right)(1-L_1-L_2)^m
\end{equation*}
where $p_l = L(\rho)$ is the state leakage of the initial state.
\begin{proof}See \cref{proof:dle-leak} \end{proof}
\end{lemma}

Notice that the leakage accumulation model in \cref{prop:dle-leak} is independent of the reduced dynamics of the map $\2E_1$ on the computational subspace. It \emph{only} depends on the leakage and seepage rates of $\2E_L$. Furthermore, since the leakage in the DLE is depolarizing, and hence purely incoherent, the coherence of leakage of the output state is always zero.

\subsubsection{Depolarizing Leakage Model}
\label{sec:depol-leakage}

An important subset of DLE channels is the case where computational subspace component $\2E_1$ is itself a depolarizing channel. We call these types of channels \emph{depolarizing leakage models} (DLMs) and they are describe by the channel 
\begin{equation}
\begin{aligned}
\2E_{D} =& (1-L_1) \Big(  \mu_1 \2I_1 + (1-\mu_1) \2D_{1} \Big) 
		\\&+ L_1 \2D_{21} + L_2\2D_{12} + (1-L_2)\2D_{2}.
\end{aligned}
\label{def:dle}
\end{equation}
where $1-\mu_1$ is the depolarizing probability of the $\2E_1$ component.

A DLM can be constructed from an arbitrary channel by performing a twirl over the computational subspace, while depolarizing the leakage subspace. Consider two independent chosen unitaries $U_1\in \1C_1$, $V_2 \in \1P_2$ where $\1C_1$ is a set of gates on the computational subspace which form a unitary 2-design, and $\1P_2$ is a set of gates on the leakage subspace which forms a unitary 1-design. The DLE projection of an arbitrary CPTP map $\2E$ is given by the independent average over both these groups:
\begin{align}
\2E_D
=&
\frac{1}{|\1C_1||\1P_2|} \sum_{\2U_1\in \1C_1}\sum_{\2U_2,\2V_2\in \1P_2}(\2U_1+\2V_2)\circ \2E\circ (\2U^\dagger_1+\2U_2)\\
=& \2W_1(\2E) + \2D_1\2E\2D_2 + \2D_2 \2E \2D_1 + \2D_2 \2E\2D_2
\label{eq:dlm-proj}
\end{align}
where $\2D_j$ is the completely depolarizing channels on $D(\2X_j)$, and $\2W_1$ be the twirling superchannel acting on the computational subspace as:
\begin{align}
\2W_1(\2E_{11}) &= \mu_1 \2I_1 + (1-\mu_1) \2D_{11} \\
\mu_1 	&= \frac{d_1\overline{F}(\2E_{11})-1}{d_1-1}. 
\label{eq:twirl-param}
\end{align}
The utility of twirling in this manner is that the resulting channel $\2E_D$ will have the same average gate fidelity, leakage rate, and seepage rate of the original channel $\2E$.
\begin{align}
\overline{F}(\2E_D) =& \overline{F}(\2E)\\
L_1(\2E_D) =& L_1(\2E)\\
L_2(\2E_D) =& L_2(\2E).
\end{align}

One advantage of considering a DLM instead of a general DLE is that we can derive a simple expression for the fidelity decay model for repeated applications of a DLM. This is the decay model used for LRB, though in the absence of SPAM errors:
\begin{align}
\overline{F}(\2E_D^m)
&= \frac{1}{d_1}\Big[ 1-p_l(m) 
	+ (d_1-1)(1-p_1)^m\mu_1^m\Big]
\label{prop:dlm-fid}
\end{align}
where $p_l(m) = L(\2E_D^m)$ is the DLE leakage accumulation model given in \cref{prop:dle-leak}.
While relatively simple, the DLM if of practical interest as it is the ideal model that LRB attempts to twirl an arbitrary error channel into.

\subsection{Unitary Leakage Model}
\label{sec:unitary-leakage}

While the DLE and DLM are useful logical models for considering leakage errors in quantum gates we can consider more specific error models based on the control Hamiltonian used to generate the quantum gate. Unitary leakage is generated by a Hamiltonian term which couples states in the computational subspace and leakage subspace. The simplest such case is generated by an exchange interaction between a state in the computational subspace $(\ket1)$ and a state in the leakage subspace $(\ket2)$. In this case the interaction Hamiltonian is given by
\begin{equation}
H = \frac12\Big(\ketbra12 + \ketbra21\Big),
\end{equation}
and the resulting unitary leakage error after evolving under $H$ for time $t$ is given by
\begin{align}
U = e^{-i t H} &= \id + \Big(\cos(t/2)-1\Big)\Big(\ketbra11 + \ketbra22\Big) 
	\nonumber\\&
	\qquad+ \sin(t/2) \Big(\ketbra12 + \ketbra21\Big). 
\label{eq:unitary-leakage}
\end{align}
In this case the leakage rate of the channel, and the state leakage of an initial state $\rho_1$, as a function of evolution time are given by
\begin{align}
L(\rho_1(t)) =& \sin^2\left(\frac{t}{2}\right)  \bra{1}\rho_1\ket{1}\\
L_j(\2U(t)) =& \frac{1}{d_j}\sin^2\left(\frac{t}{2}\right), \quad j=1,2. \label{eq:unitary-L12}
\end{align}
As mentioned in \cref{sec:leakage-errors} we have that the leakage and seepage rates satisfy $d_2L_2 = d_1 L_1$. 

We find that for this interaction Hamiltonian the state leakage oscillates as a function of time, which is a distinct difference from the DLE leakage accumulation model in \cref{prop:dle-leak}. This is because the leakage error is generating coherent Rabi-oscillations between a state in the computational and in the leakage subspaces. Accordingly, this type of leakage also generates coherences in the leakage if the initial state has some population in $\ket{1}$ or $\ket{2}$. For example, if the initial sate is $\rho_1(0) = \ketbra11$ we have that the coherence of leakage at time $t$ is given by
 \begin{equation}
CL( \rho_1(t))= | \sin(t) |.
\end{equation}
Further more, the coherent leakage rate at time $t$ is given by
\begin{align}
CL_1(\2U(t))=&\frac{2 \left| \sin \left(\frac{t}{2}\right)\right|  \Big[2-\big(1+\cos (t)\big) \left| \cos \left(\frac{t}{2}\right)\right|\Big]}{3 \big(1-\cos (t)\big)}
\end{align}
and the upper-bound for $CL_1$ from \cref{prop:coh-leakage-bound} is 
\begin{equation}
CL_1(\2U(t)) 
	\le \frac{2}{d_1}\left|\sin\left(\frac{t}{2}\right)\right|\sqrt{d_1-\sin^2\left(\frac{t}{2}\right)}.
\end{equation}

\subsubsection{DLM of a Unitary Leakage Error}
\label{sec:unitary-dlm}

We can also consider the perfect depolarizing projection of the unitary error onto a DLM after a fixed time $\Delta t$. In this case from inserting the expressions for $L_j$ from \cref{eq:unitary-L12} into the leakage accumulation model in \cref{prop:dle-leak} we have
\begin{align*}
p_l(m) 	=&  \frac{d_2}{d_1+d_2} - \frac{d_2}{d_1+d_2}\left[1-\frac{d_1+d_2}{d_2d_1}\sin^2\left(\frac{\Delta t}{2}\right)\right]^m	
\end{align*}
If we suppose we have a qutrit leakage model ($d_1=2,d_2=1$) then this reduces to
\begin{equation}
p_l(m) 	=  \frac13 - \frac13\left[1-\frac32\sin^2\left(\frac{\Delta t}{2}\right)\right]^m. \label{eq:qutrit-u-leak}
\end{equation}
Note that after projection onto a DLM this will no longer generate oscillations.

\begin{figure}[h]
\centering
\includegraphics[width=0.95\columnwidth]{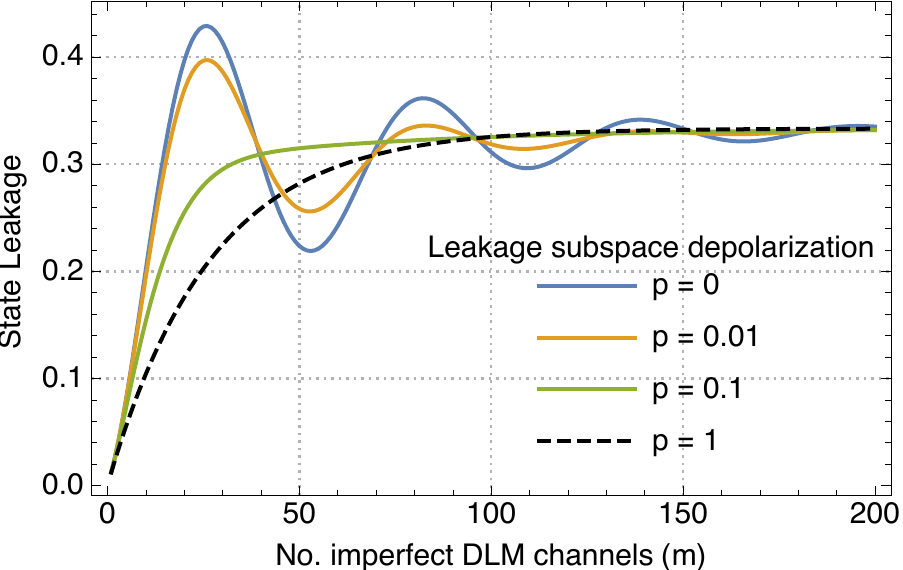}
\caption{State leakage accumulation of the imperfect DLM projection of the unitary leakage error channel in \cref{eq:unitary-leakage} allowing partial coherences between the leakage and computational subspaces. This results in memory effects in the leakage subspace which may give rise to the observed oscillations in state leakage.  The black dotted curve is the ideal case of a perfect DLM projection corresponding to an exponential leakage accumulation model.}
\label{fig:coherent-unitary}
\end{figure}

What happens if we have an imperfect projection due to not perfectly implementing the required twirling procedure in \cref{eq:dlm-proj}? In this case coherences between the leakage and computational subspaces may survive and be observed as memory effects resulting in oscillations in the resulting leakage accumulation model. We consider this by computing a twirl over the Clifford group on the computational subspace, where each perfect Clifford gate is extended to act as the identity on the leakage subspace $U = U_1 \oplus \id_2$. This is followed by an imperfect depolarizing of the leakage subspace by a depolarizing channel with depolarizing probability $p$
\begin{equation}
\2D(p) = (1-p) \2I + p (\2I_1 + \2D_2),
\end{equation}
 where $\2I$ is the identity channel on the full Hilbert space, and $\2I_1, \2D_2$ are the identity channel, and completely depolarizing channel on the computational and leakage subspaces respectively. The leakage accumulation due to repeated applications of the resulting imperfectly twirled DLM for different values of depolarizing strength $p$ are shown in \cref{fig:coherent-unitary}. We find here that when there is no depolarizing of the leakage subspace that coherent oscillations survive for some time before being damped out. For a depolarizing strength of 10\% these oscillations are damped out, however we still observe faster leakage accumulation than the completely depolarized ideal case which would lead to an overestimate of the leakage rate in $L_1 = \frac12 \sin^2(\Delta t/2)$ in \cref{eq:qutrit-u-leak}.

\subsubsection{Weak Unitary Leakage}

Now let us consider a general a unitary error model which is more applicable to many experimental scenarios where unitary leakage is introduced into a system by imperfections in a control hamiltonian. In general the leakage due to a unitary term is give by
\begin{equation}
U(t) = \2T\exp\left(-i \int_0^t dt_1 H(t_1)\right)
\end{equation}
where $\2T$ is the time-ordering operator, and $H(t)$ is the time-dependent Hamiltonian the system evolves under. The leakage and seepage rates of this interaction are given by
\begin{align}
L_j(\2U(t)) =& \frac{1}{d_j}\Tr[\id_2  U(t) \id_1 U(t)^\dagger],\quad j=1,2.
\end{align}

By expanding $U(t)$ into a Dyson series, we may obtain a perturbation expansion for the leakage and seepage rates due to the unitary leakage error. In doing so we find that the second order Dyson term is the leading order contribution to leakage and seepage rates, and is given by
\begin{equation}
L_j(\2U(t)) \approx  \frac{t^2}{d_j} \Tr[\id_2 \overline{H}(t) \id_1 \overline{H}(t)]
\end{equation}
where $\overline{H}(t) =  \frac{1}{t}\int_0^t dt_1 H(t_1)$ is the first order average Hamiltonian term of $H(t)$. Hence we can estimate the leakage rates due to given control pulse by computing the average Hamiltonian over the pulse shape. 

\begin{figure}[h]
\centering
\includegraphics[width=0.95\columnwidth]{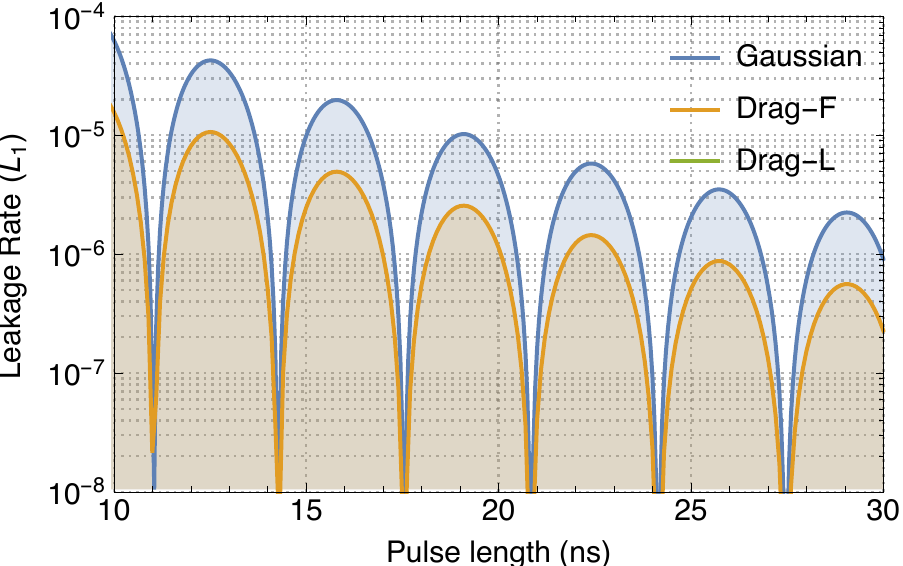}
\caption{First order unitary leakage rate for an $X-\pi/2$ rotation pulse for a transmon qubit with anharmonicity $\delta/2\pi = -300$~MHz. The Drag-L pulse is not visible on the log plot as the first order leakage rate is 0.}
\label{fig:weak-unitary}
\end{figure}

To illustrate this consider the transmon qubit system as used in \cref{sec:lrb-sim}. The first order leakage rate contribution from a $X_{\pi/2}$ control pulse is given in \cref{fig:weak-unitary}. Here we compare the same DRAG pulse shapes used in \cref{eq:drag} with DRAG parameters $\alpha=0$ (Gaussian), $\alpha=0.5$ (Drag-F), and $\alpha=1$ (Drag-L) for a transmon with anharmonicity $\delta/2\pi = -300$~MHz. For the Drag-L pulse the first-order unitary leakage is 0.

\subsection{Lindblad Leakage Models}
\label{sec:lindblad-leakage}

 In the open quantum systems framework CPTP maps are generated by exponentiation of a Lindblad generator $\2E = e^{t(\2H+\2D)}$ where $\2H$ and $\2D$ are the generators of purely unitary and purely dissipative evolution respectively:
\begin{align}
\2H(\rho) =& -i [H,\rho] \label{eq:lind-unitary}\\
\2D(\rho) =& \sum_k \gamma_k  A_k\rho A_k^\dagger -\frac12\{A_k^\dagger A_k,\rho\}. \label{eq:lind-diss}
\end{align}

In a real experiment leakage will not generally be purely dissipative or purely unitary, but a combination of both, however for calibration and gate optimization it is useful to estimate the relative contributions from both the dissipative and unitary parts individually. In practice, the dissipative leakage contribution will be an \emph{always-on} process that is due to thermal relaxation or other incoherent interactions, and is typically beyond the experimenters direct control. The unitary contribution, however, will typically be due to control errors which may be optimized, or interaction terms with neighboring systems which may be decoupled via control. Estimation of these two quantities may be achieved by considering a short-time expansion of $\2E$. This is useful, as for many commonly used models of dissipation the leakage contributions from unitary and dissipative processes are additive up to second order, which we prove in the following proposition.

\begin{lemma} \label{prop:lindblad-leakage}
Let $\2E$ be a CPTP channel with Lindblad generators $\2E=\exp\left(\Delta t(\2H +\2D)\right)$, where the dissipation operators $A_k$ are all raising or lowering operators:
\[
A_{\pm k} = \sum_j \alpha_j \ketbra{j\pm k}{j} 
\]
To second order in $\Delta t$ we have that the leakage and seepage rates for $\2E$ are given by
\begin{equation}
L_j(\2E) = L_j(\2E_{\text{uni}})  +L_j(\2E_{\text{diss}}) \,\quad j=1,2
\end{equation}
where $\2E_{\text{uni}}$ and $\2E_{\text{diss}}$ are purely unitary and purely dissipative CPTP maps generated by $\2H$ and $\2D$ respectively.
\begin{proof}
See \cref{proof:lindblad-leakage}
\end{proof}
\end{lemma}

One could use \cref{prop:lindblad-leakage} as a coarse way to estimate the contribution of a unitary leakage error from a LRB experiment in the presence of an always-on thermal leakage error which can general have the relevant dissipation parameters measured independently. If we return to the LRB simulation in \cref{sec:lrb-sim}, for example, dissipative effects were due to thermal $T_1$ relaxation. Using estimated values for $T_1$ and the average photon number of the system we may compute the theoretical dissipative contribution under an appropriate relaxation model and subtract them from the LRB error estimates to obtain coarse estimates for the unitary contribution. We now give some explicit examples of phenomenological dissipation models which may generate leakage.

\subsubsection{Simple Dissipative Leakage}
\label{sec:diss-leakage}

Consider the CPTP map generated by a purely dissipative leakage model that couples a single state in the computational subspace ($\ket1$) with a single state the leakage subspace ($\ket2$). In this case our Lindblad dissipator consists of two operator terms
\begin{align}
A_{21} =& \ketbra21,\quad&
A_{12} =& \ketbra12
\end{align}
with corresponding rates $\gamma_1,\gamma_2$. This first term generates leakage  from the $\ket1$ state to the $\ket2$ state at a rate $\gamma_1$, while the second term generates seepage from $\ket2$ to $\ket1$ at a rate $\gamma_2$. For this simple case the resulting error channel, given by the superoperator in $\2S(t) = e^{t \2D}$, can be evaluated analytically. The leakage and seepage rates for this map as a function of time are given by 
\begin{align}
L_1(\2E) =& \frac{\gamma_1}{d_1(\gamma_1+\gamma_2)} \left(1-e^{-t(\gamma_1+\gamma_2)}\right) \\
L_2(\2E) =& \frac{\gamma_2}{d_2(\gamma_1+\gamma_2)} \left(1-e^{-t(\gamma_1+\gamma_2)}\right).
\end{align}
If we have an initial state $\rho$, then the state leakage as a function of time is given by
\begin{equation}
\begin{aligned}
L(\rho(t)) 
	=& L\big(\rho(0)\big)+\\&
	\frac{\gamma_1\bra{1}\rho\ket{1} - \gamma_2\bra{2}\rho\ket{2}}{
	\gamma_1+\gamma_2}\left[1- e^{-t(\gamma_1+\gamma_2)}\right].
\end{aligned}
\end{equation}

We can also consider more general dissipation models, however in order to compute the leakage rates in these cases we will generally have to consider a short time expansion of the dissipative superoperator.

\subsubsection{Thermal Relaxation}
\label{sec:cavity-leakage}

Let us now consider a physically motivated example of leakage due to thermal relaxation in a harmonic, or weakly anharmonic, oscillator. In such a system thermal relaxation to an equilibrium state is described by the Markovian photon-loss dissipator~\cite{Wiseman2009book}
\begin{align}
\2D_c &= \gamma_\downarrow D[a] + \gamma_\uparrow D[a^\dagger] \\
 &= \kappa(1+\overline{n}) D[a] +\kappa \overline{n} D[a^\dagger]
\label{eq:cavity-diss}
\end{align}
where we have defined the dissipation rate $\kappa = \gamma_\downarrow - \gamma_\uparrow \ge 0$, and the \emph{average photon number} of the oscillator
\begin{equation}
\overline{n} = \frac{\gamma_\uparrow}{\gamma_\downarrow-\gamma_\uparrow}.
\end{equation}
The state leakage of the cavity at thermal equilibrium by
\begin{align}
L(\rho_{eq})=& \left(\frac{\gamma_\uparrow}{\gamma_\downarrow}\right)^2
  = \left(\frac{\overline{n}}{1+\overline{n}}\right)^2.
\end{align}
If we consider the error channel for evolution over a time $\Delta t$ such that $\kappa \Delta t \ll 1$, then to second order we find that the leakage and seepage rates are given by 
\begin{align}
L_1(\2E) \approx& \kappa \overline{n}\,\Delta t\left[1 - (3+4\overline{n}) \kappa \Delta t\right] \\
L_2(\2E) \approx& \frac{2(1+\overline{n})\kappa \Delta t}{d_2}\left[1+(1-4\overline{n})\kappa \Delta t \right].
\end{align}

\begin{figure}[h]
\centering
\includegraphics[width=\columnwidth]{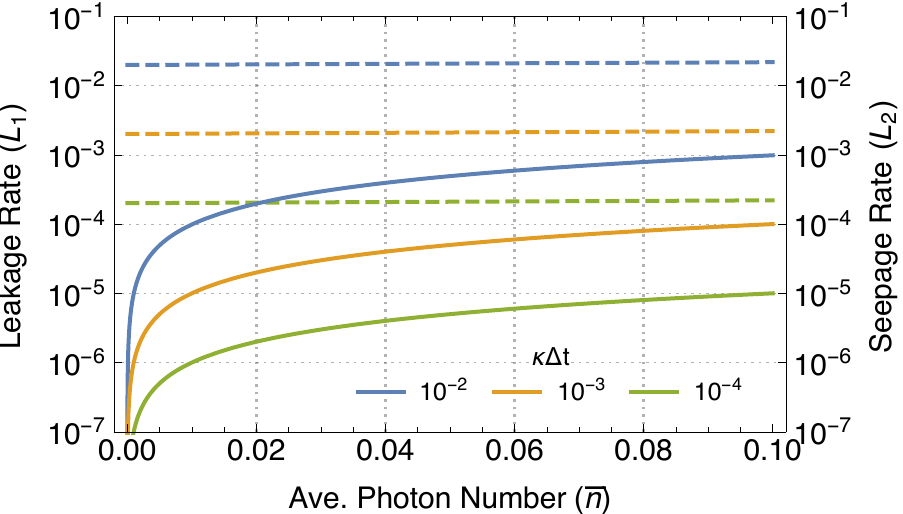}
\caption{Plot of leakage (solid) and seepage (dashed) rates vs average photon number at thermal equilibrium for $T_1$ relaxation of a qubit encoded in the lowest two energy levels of a cavity. The leakage rate is much less than the seepage rate $(L_1 \ll L_2)$ across the plotted photon number range.}
\label{fig:cavity-leakage}
\end{figure}

Hence in the low photon limit ($\overline{n}\ll1$) we have that $L_2(\2E) \gg L_1(\2E)$. This is illustrated in \cref{fig:cavity-leakage} where we plot the leakage and seepage rate vs equilibrium photon number for $\overline{n}=0$ to $0.1$ for values of $\kappa \Delta t = 10^{-4}, 10^{-3}, 10^{-2}$. 

Note that for a true cavity $d_2 = \infty$, so we must truncate the dimension of the cavity to some reasonable number of excitations. In the low $\overline{n}$ limit we may truncate to a qutrit model ($d_2=1$). The rates shown in \cref{fig:cavity-leakage} are comparable to those expected for a superconducting transmon qubit which typically have average photon numbers in the range of $\overline{n}\approx 10^{-2}$ to $10^{-1}$.

\subsection{Multiple Leakage Subspaces}
\label{sec:multi-leakage}

The leakage errors described in \cref{sec:leakage-errors} report an average rate for leakage between the computational subspace and the \emph{entire} leakage subspace. If the computational subspace corresponds to a composite system, for example an $n$-qubit system, there may be several different leakage rates to different levels in the leakage subspace corresponding to leakage of each individual system, or cross-system leakage across components. For example in superconducting qubit systems there may be multiple different leakage rates due to frequency crowding in the off-resonant leakage levels and cross-talk in system control~\cite{Sheldon2016pra2,Takita2016prl}. In such situations useful characterization may require a more fine-grained approach.

For composite-systems the definitions for state leakage, leakage rates, and seepage rates defined in \cref{sec:leakage-errors} naturally generalize to describe leakage to multiple leakage subspaces by simply decomposing the leakage subspace into direct sum subspaces 
\begin{equation}
\2X_2 = \2Y_{1} \oplus \hdots \oplus \2Y_{m}.
\end{equation}
Using this decomposition we may define $m$ different measures of state leakage, leakage rates, and seepage rates which are given by replacing the projector $\id_2$ with the projector $\id_{2_{\2Y_j}}$ onto $\2Y_j$ in \cref{eq:state-leakage,eq:leakage-rate}:
\begin{align}
L_{\2Y_j}(\rho) =& \Tr[\id_{\2Y_j}\rho] \\
L_{1_{\2Y_j}}(\2E) =& \Tr\left[\id_{\2Y_j} \2E\left(\frac{\id_1}{d_1}\right)\right]\\
L_{2_{\2Y_j}}(\2E) =& \Tr\left[\id_1 \2E\left(\frac{\id_{\2Y_j}}{d_{\2Y_j}}\right)\right]
\end{align}
where $d_{\2Y_j}$ is the dimension of $\2Y_j$, and $j=1,\hdots m$.

The definitions of the total state leakage, leakage rate, and seepage rate the full leakage subspace may be expressed in terms of these multi-rate definitions as 
\begin{align}
L(\rho) &= \sum_{j=1}^m L_{\2Y_j}(\rho)	\\
L_1(\2E) &= \sum_{j=1}^m L_{1_{\2Y_j}}(\2E) \\
L_2(\2E) &= \sum_{j=1}^m d_{2_j}L_{2_{\2Y_j}}(\2E).
\end{align}

Let us consider a simple example of 2-qubit leakage where we define three leakage subspaces corresponding to only the first qubit leaking, only the second qubit leaking, and both qubits leaking. In this case our computational subspace is given by the tensor product of the computational subspaces of each of the qubits: $\id_{\2X_1} = \id_1\otimes\id_1$. The leakage subspaces are then given by
\begin{align}
\id_{\2Y_1} = \id_2\otimes\id_1 \\
\id_{\2Y_2} = \id_1\otimes\id_2 \\
\id_{\2Y_3} = \id_2\otimes\id_2
\end{align}
and hence the full leakage subspace is $\2X_2 = \id\otimes\id - \2X_1$.

In the most general case we could have that each of the $\2Y_1$ correspond to a 1-dimensional subspace spanned by one of the leakage basis states of $\2X_2$. Note that under this assumption we are ignoring direct interactions between individual leakage subspaces --- any interacting subspaces in this sense should be considered as a single subspace. An important direction for future research is to develop characterization methods for these multi-qubit system with multiple leakage subspaces.

\section{Conclusion}
\label{sec:conc}

We have presented a framework for the quantification of leakage errors, both coherent and incoherent, in quantum systems and a method for characterizing the average properties of leakage errors in a quantum gate set by leakage randomized benchmarking. These tools provide a means for evaluation of new methods for suppressing and correcting leakage errors in quantum systems. 

An important point in characterizing leakage errors in quantum gates is that two rates are required to specify average leakage dynamics, the leakage rate out of the computational subspace, and the seepage rate back into the computational subspace. We illustrated this with several examples demonstrating leakage mechanisms due to control errors and thermal relaxation processes. Further, leakage rates themselves are not necessarily predictive of the average performance of quantum gates as specified by the average gate fidelity. This is because leakage dynamics can induce logical errors within the computational subspace. This has been well documented, and typically results in a phase error due to population briefly spending time in an off-resonant leakage level before returning to the computational subspace by the end of a control pulse. This can be seen explicitly in our simulation of leakage randomized benchmarking for a superconducting qubit where we contrasted Gaussian control pulses with two types of DRAG pulses designed to correct the phase-error, and to suppress leakage rates, and has also been studied in recent experimental works \cite{Chen2016prl,McKay2016arx}. 

The LRB protocol we presented provides a theoretical justification for the technique first used in Ref. \cite{Chen2016prl}, and in particular we find that the key assumption for the validity of the decay model is that twirling the computational subspace also acts to completely depolarize the leakage subspace. If this assumption breaks down, then in the presence of coherent leakage errors, such as those from unitary dynamics, non-Markovian effects may manifest as oscillations in the LRB decay curves about the ideal exponential model. In the case of the simple example we considered, even though these oscillations are quickly damped out by partial depolarization, they could lead to an overestimate of the leakage and seepage rates of the system.

While we focused on RB based characterization methods in the present article we comment briefly on full tomographic methods for gate characterization. Under the assumption that one cannot directly measure leakage levels one cannot fully characterize average leakage dynamics by quantum process tomography. This is because at best one can only reconstruct the computational subspace channel component, which in the presence of leakage is not a trace-preserving map. From this channel component the leakage rate could in principle be determined by the sub-normalized trace of the reconstructed channel, but it is not obvious how to estimate the seepage rates. To estimate leakage and seepage in a tomographic setting one would have to use a sequence of gates amplify a leakage decay model. This could be done using recent additions to the gate-set tomography protocol which use germ sequences of increasing length to amplify errors~\cite{Dehollain2016njp}. By combining gate-set tomography with the depolarizing leakage extension channel we developed in \cref{sec:depol-leakage} one could attempt to reconstruct the effective channel on the computational subspace along with the leakage and seepage rates in the case where GST seeds also act to implement a depolarizing channel on the leakage subspace.

\acknowledgements

We thank 
D. McKay for helpful discussion and suggestions. This work was supported by ARO under contract W911NF-14-1-0124.

\bibliography{cjwood-leakage}

\section*{Appendices}
\begin{appendix}

\section{Derivation of the LRB Decay Model}
\label{sec:lrb-decay-model}

Consider a leakage system with state space $\2X= \2X_1\oplus\2X_2$, and a unitary operator $U_1\in L(\2X_1)$ that acts on the computational subspace $\2X_1$. To extend $U_1$ to a unitary on $L(\2X)$ we may add a unitary $U_2\in L(\2X_2)$ such that $U_{12}\equiv U_1\oplus U_2 $ is unitary. Ideally, up to a phase this target extension will be the identity operator ($U_2=\id_2$), so that our intended interaction on $\2X_1$ acts trivially on $\2X_2$. 

Let $\2U_{12}$ represent the quantum channel for unitary evolution $\2U_{12}(\rho)\equiv U_{12}\rho U_{12}^\dagger$. The superoperator representation of $\2U_{12}$ in the column-stacking convention is given by
\[
\2S_{\2U_{12}} 
= (U_1\oplus U_2)^*\otimes(U_1 \oplus U_2) 
= \2S_{\2U_1} +\2S_{\2U_2} + \2S_{\delta_{12}}, 
\]
where we
$\2S_{\2U_1} =  (U_1^*\oplus 0)\otimes (U_j\oplus 0)$ is a unitary superoperator with support on $\2L(\2X_1)$, and similarly for $\2S_{\2U_2}$ on $L(\2X_2)$, and 
\begin{equation}
\2S_{\delta_{12}}= (U_1\oplus 0)^*\otimes(0\oplus U_2) + (0\oplus U_2^*)\otimes (U_1\oplus 0)
\label{eq:delta-coh}
\end{equation}
is a superoperator component which acts on the CLS defined by the projector in \cref{eq:cls-proj} of the main text. 

Let $\2C_k$ be the noisy implementation of the extension $\2U_{12,k}$ of the Clifford matrix $\2U_{1,k}$ on the computational subspace:
\begin{align}
\2C_k =& \2E \circ \2U_{12,k}\\
\2S_{\2C_k} =& \2S_{\2E} \circ (\2S_{\2U_{1,k}} + \2S_{\2U_{2,k}}+ \2S_{\delta_{12}}).
\end{align}
\begin{assumption} 
Here we have assumed a zeroth order approximation where the noise channel is approximately constant as a function of time, and for each Clifford gate $\2C_k$. 
\end{assumption}
As with the case of standard randomized benchmarking the decay model derived in this limit should be valid even for slightly gate dependent noise~\cite{Magesan2012pra}. 

Consider now the randomized benchmarking protocol of choosing a sequence 
\[
i_m = \2C_m \circ \hdots \circ \2C_1
\]
of $m$ Clifford gates, where the order of composition is such that $\2C_1$ is applied to the system first. The $m+1$ gate is chose to be the usual recover operation 
\[
\2C_{m+1} = \2E\circ \2R_{i_m}
\] 
where on the computational subspace we have that the recovery operator satisfies
\[
\2R_{1,i_m} = \2U_{1,1}^\dagger \circ \hdots \circ \2U_{1,m}^\dagger.
\]
The full RB sequence is then given by $i_m^\prime = \2C_{m+1}\circ i_m$.

\begin{assumption}
To further evaluate this sequence and evaluate the average leakage and seepage rates of the noisy gateset we want to project $\2U_{12}$ onto the ILS defined in \cref{eq:ils-proj} so that $\2S_{\delta_{12}} \approx 0$, and 
\begin{equation}
\2S_{\2U_{12,k}} \approx S_{\2U_{1,k}} +\2S_{\2U_{2,k}}.
\label{eq:u-coherence-cancel}
\end{equation}
\end{assumption}

One method of doing this is to consider averaging over a local phase on one of the subspaces as was used in previous work \cite{Epstein2014pra,Chasseur2015pra,Wallman2016njp}. If one can implement the superoperator for $U_{-12} = -U_1 + U_2$ with a negative local-phase on $\2U_1$ (or equivalently on $\2U_2$), then the average of the two superoperators is given by
\begin{equation}
\2S_{\overline{\2U}_{12}} \equiv \frac12\left(\2S_{\2U_{12}} +\2S_{\2U_{-12}}  \right) = \2S_{\2U_1} +\2S_{\2U_2}.
\end{equation}
We note that it may be difficult to experimentally implement this local phase difference without control of the leakage subspace, however, in practice it appears to make little difference for weak leakage rates demonstrated in \cite{Chen2016prl,McKay2016arx}. 

Using the assumption in \cref{eq:u-coherence-cancel} the superoperator for the RB sequence $i_m^\prime$ is given by
\begin{align}
\2S_{i_m^\prime} 
	&= \2S_{\2E} (\2S_{\2R_{1,i_m}} + \2S_{\2U_{2,m+1}}) \hdots \2S_{\2E} (\2S_{\2U_{1,1}} + \2S_{\2U_{2,1}}).
\end{align}
The  survival probability for an initial state $\rho$, and measurement of an operator $M$ for a gate sequence $i_m^\prime$ is given by
\begin{align}
P(1 | i_m^\prime\, M,\rho) &= \dbra{M}\2S_{i_m^\prime}\dket{\rho} = \Tr[M^\dagger \,\2S_{i_m^\prime}(\rho)]
\label{eq:rb-survival-seq}
\end{align}
For a given length $m$ sequence of Clifford gates, the target decay model for randomized benchmarking is given by the average of \cref{eq:rb-survival-seq} over all equal  sequences $i_m^\prime$:
\begin{align}
P(1 | m, M,\rho) &\equiv \1E_{i_m^\prime}[P(1 | i_m^\prime\, M,\rho) ] \nonumber\\
	&= \dbra{M}\1E_{i_m^\prime}[\2 S_{i_m^\prime}]\dket{\rho}.
	\label{app-eq:survival}
\end{align}

To evaluate \cref{app-eq:survival} we may express $i_m^\prime$ in terms of unitaries $\2V_{j,k} = \2U_{j,1}^\dagger \hdots \2U_{j,k}^\dagger$ so that
\begin{align}
\2S_{i_m^\prime} 
	&= \2S_{\2E} (\2S_{\2I_1} + \2S^\dagger_{\2V_{2,m+1}}) \times\nonumber\\&\qquad\prod_{k=1}^m \Big[
	(\2S_{\2V_{1,k}} + \2S_{\2V_{2,k}})\2S_{\2E}
	(\2S_{\2V_{1,k}}^\dagger + \2S_{\2V_{2,k}}^\dagger)
	\Big]. \label{eq:s-rb-seq}
\end{align}
\begin{assumption}
To proceed we must make another assumption that we may \emph{independently average} over the sets $\{U_{1,k}\}$ and $\{U_{2,k}\}$, we have that
\begin{align}
\1E_{i_m^\prime}[\2 S_{i_m^\prime}]
	&=  \2S_{\2E} \left(\2S_{\2I_1} 
		+ \1E_{i_m^\prime} \left[ \2S^\dagger_{\2V_{2,m+1}} \right] \right) 
		\label{eq:s-rb-average}
		\cdot\\&\quad 
	\left( \1E_{i_m^\prime} \left[
	(\2S_{\2V_{1,k}} + \2S_{\2V_{2,k}})\2S_{\2E}
	(\2S_{\2V_{1,k}}^\dagger + \2S_{\2V_{2,k}}^\dagger)
	\right] \right)^m. \nonumber
\end{align}
\end{assumption}
Now, since the Clifford group $\{\2U_{1,k}\}$ is a unitary 2-design we make use of the twirling identity
\begin{align}
\1E_{i_m^\prime} \left[\2S_{\2V_{1,k}} \2S_{\2E} \2S_{\2V_{1,k}}^\dagger  \right] &= \2W_1(\2E)
\end{align}
where 
\begin{align}
\2W_1(\2E) &= \mu_1 \2I_1 + (1-\mu_1) \2D_{1} \\
\mu_1 	&= \frac{d_1\overline{F}_{\2E_{11}}-1}{d_1-1} \label{eq:twirl-param}
\end{align}
is the twirling superchannel acting on the computational subspace, $\2I_1$ is the identity projector on the computational subspace, and
\begin{equation}
\2D_{1}(\rho) = \Tr[\id_1 \rho] \,\frac{\id_1}{d_1}
\end{equation}
is the completely depolarizing channel on the computational subspace. Since the Clifford group is also a unitary 1-design we may also evaluate
\begin{align}
\1E_{i_m^\prime} \left[\2S_{\2V_{1,k}}\right] = \2S_{\2D_{1}}.
\end{align}

\begin{assumption}\label{assum4} We now need to make one final assumption, that the matrices $\{\2U_{2,k}\}$ averaged over leakage subspace also form a unitary 1-design to ensure that we also have
\begin{equation}
\1E_{i_m^\prime} \left[\2S_{\2V_{2,k}} \right] = \2S_{\2D_{2}}. 
\label{eq:leakage-1d}
\end{equation}
where $\2D_{2}$ is the completely depolarizing channel on the leakage subspace.
\end{assumption}
Under \cref{assum4} we have that average superoperator in \cref{eq:s-rb-average} evaluates to
\begin{align}
\1E_{i_m^\prime}[\2S_{i_m^\prime} ]
	&= \2S_{\2E}  \Big[
	\2S_{\2W_1(\2E)}
	+ \2S_{\2D_{1}} \2S_{\2E} \2S_{\2D_{2}}
	+ \2S_{\2D_{1}} \2S_{\2E} \2S_{\2D_{2}}
	\nonumber\\&\qquad+ \2S_{\2D_{2}} \2S_{\2E} \2S_{\2D_{2}}
	\Big]^m 
	\nonumber\\&=\2S_{\2E}\2S_{\2E_D}^m
\end{align}
where $\2E_D$ is given by
\begin{align}
\2E_D &= (1-L_1) \Big(  \mu_1 \2I_1 + (1-\mu_1) \2D_{1} \Big) 
	+ L_1 \2D_{21} 
	\nonumber\\&\qquad
	+ L_2\2D_{12} + (1-L_2) \2D_{2}
	\label{app-eq:dlm}
\end{align}
and $L_1, L_2$ are the leakage and seepage rate of $\2E$ respectively, and we have defined completely depolarizing channels between the computational and leakage subspace:
\begin{equation}
\2D_{ij}(\rho) = \Tr[\id_j \rho] \,\frac{\id_i}{d_i}, \quad i,j =1,2.
\end{equation}

To compute $\2S_{\2E_{D}}^m$ we let ${A_j/\sqrt{d_1} : j=0,...,d_1-1}$ be an orthonormal operator basis for $\2L(\2X_1)$, with $A_0 = \id_1$. In the qubit case this could be the Pauli basis $\{\id_1, X, Y, Z\}/\sqrt{2}$, for example. The superoperator representations of the identity and completely depolarizing channels may be expressed in this basis as
\begin{equation*}
\2S_{\2I_1} = \sum_{j=0}^{d_1-1} \frac{1}{d_1} \dketdbra{A_j}{A_j}, \qquad
\2S_{\2D_{1}} = \frac{1}{d_1} \dketdbra{A_0}{A_0}.
\end{equation*}
Hence  \cref{app-eq:dlm} may be written as $\2E_D^m = \2E_{L}^m + \2E_F^m$ where
\begin{align*}
\2E_F =& (1-L_1) \mu_1 (\2I_1-\2D_1)\\ 
\2E_L =&(1-L_1) \2D_{11} + L_1\2D_{21} + L_2 \2D_{12} + (1-L_2) \2D_{22} 
 \end{align*}
 and we have used the fact that $\Tr[\2S_{\2E_L}^\dagger \2S_{\2E_F}]=0$ to expand $(\2E_L + \2E_F)^m =\2E_L^m + \2E_F^m$.

Since the operators $\2D_{ij}$ are mutually orthogonal we can compute the exponential of the superoperator for $\2E_L$ as a $2\times 2$ matrix
\begin{align*}
\2S_{\2E_L}^m 
	&=\begin{pmatrix} 1-L_1 & L_1 \\ L_2 & 1-L_2\end{pmatrix}^m \\
	=& \frac{1}{L_1+L_2}
	\begin{pmatrix} L_2 & L_1 \\ L_2 & L_1 \end{pmatrix}
	+\frac{(1-L_1-L_2)^m}{L_1+L_2}\begin{pmatrix} L_1 & -L_1 \\ -L_2 & L_2 \end{pmatrix}
\end{align*}
and hence for an initial state $\rho$ with $L(\rho)=p_l$ we have
\begin{align}
\2S_{\2E_L}^m (\rho)
	=&  \left(\frac{L_2}{L_1+L_2}\right) \frac{\id_1}{d_1} 
		+\left(\frac{L_1}{L_1+L_2}\right)\frac{\id_2}{d_2}
		 \label{eq:slm-rho}\\&
		+\left(\frac{L_1}{L_1+L_2}-p_l\right)(1-L_1-L_2)^m \left( \frac{\id_1}{d_1} -\frac{\id_2}{d_2}\right).  \nonumber
\end{align}
For the $\2E_F$ component we simply have
\begin{equation}
\2E_F^m = (1-L_1)^m \mu_1^m (\2I-\2D)
\end{equation}
and hence
\begin{align}
\2E_F^m(\rho) =& (1-L_1)^m \mu_1^m (1-p_l) \left(\rho_1 - \frac{\id_1}{d_1}\right) \label{eq:sfm-rho} 
\end{align}
where $\rho_1$ is defined by the projection onto the computational subspace $\2I_1(\rho) = (1-p_l)\rho_1$. 

Next, using the the survival probability in \cref{app-eq:survival} we consider outcomes for a set of measurements $\{M_j\}$ that ideally form a PVM on the computational subspace $(M_j = \ketbra{j}{j})$. Using the expressions in \cref{eq:sfm-rho,eq:slm-rho} we have
\begin{align}
P(1 |, m, M_j, \rho)
	&= A_j + B_j \lambda_1^m + C_j \lambda_2^m 
\end{align}
where
\begin{align}
\lambda_1 & = 1-L_1-L_2\\
\lambda_2 & = (1-p_1)\mu_1\\
A_j & = \frac{1}{L_1+L_2}\Tr\left[M_j^\dagger\2E\left(
			L_2 \frac{\id_1}{d_1}+L_1\frac{\id_2}{d_2}
			\right)\right]  \\
B_j & =  \left(\frac{L_1}{L_1+L_2} -p_l\right) \Tr\left[M_j^\dagger \2E\left(
		 	\frac{\id_1}{d_1}-\frac{\id_2}{d_2}
			\right)\right]\\
C_j & = (1-p_l)\Tr\left[M_j^\dagger \2E\left(\rho_1 - \frac{\id_1}{d_1}\right)\right].
\end{align}
By setting $M_0=\rho_1$ as the ideal by measurement this gives the RB fidelity decay model for $j=0$. 

For the leakage model we must sum over the survival probabilities for the set of PVM measurements $\{M_j\}$. To allow for leakage in our measurement we assume a measurement leakage model given by
\begin{equation}
\sum_j M_j = (1-q_1)\id_1 + q_2 \id_2
\end{equation}
where $q_1$ and $q_2$ are the measurement leakage and seepage rates. Using this model for measurement leakage we have $\sum_j C_j = 0$, and hence the summed decay model is given by
\begin{equation}
P(1 |, m, \id_1, \rho) = A + B \lambda_1^m 
\end{equation}
where
\begin{align}
A & = \frac{L_2}{L_1+L_2}  
	+\frac{L_1 q_2 - L_2 q_1}{L_1+L_2}  \\
B & =  
	\frac{L_1}{L_1+L_2} - \frac{L_1(q_1+q_2)}{L_1+L_2}
	-p_l (1-q_1-q_2) 
\end{align}
Let us define two error terms
\begin{align}
\epsilon_M &= q_1 + p_l (1-q_1-q_2)\\
\epsilon_Q &= L_1q_2-L_2q_1
\end{align}
Then we may rewrite $A$ and $B$ as
\begin{align}
A & = \frac{L_2+\epsilon_Q}{L_1+L_2},  &
B & =  \frac{L_1-\epsilon_Q}{L_1+L_2} -\epsilon_M 
\end{align}
Hence our estimates of $L_1,L_2$ as computed from $A$ and $B$ are given by
\begin{align*}
L_1^{\text{est}}(A) &= (1-A)(1- \lambda_1) = L_1 -\epsilon_Q\\
L_2^{\text{est}}(A) &= A(1-\lambda_1) = L_2 + \epsilon_Q\\
L_1^{\text{est}}(B) &= B(1-\lambda_1) = L_1 -\epsilon_Q -\epsilon_M(L_1+L_2)\\
L_2^{\text{est}}(B) &= (1-B)(1-\lambda_1) = L_2 +\epsilon_Q +\epsilon_M(L_1+L_2)
\end{align*}
Hence the variance due to using the approximate model is less using $A$ then $B$, and is given by 
\[
\mbox{Var}(L_{j}^{\text{est}}) = \epsilon_Q^2.
\]

\section{Proof of \cref{prop:coherence-bound}}\label{proof:coherence-bound}

Consider the spectral decomposition of a state $\rho=\sum_a\lambda_a\ketbra{\Psi_a}{\Psi_a}$. We can decompose each eigenstate $\ket{\Psi_a}$ as 
\begin{equation}
\ket{\Psi_a}= \sqrt{1-p_a}\ket{\psi_{1,a}}
	+\sqrt{p_a}\ket{\psi_{2,a}} \nonumber
\end{equation}
where $0\le p_a\le 1$ is the leakage of the state $\ket{\Psi_a}$. The leakage of $\rho$ is then given by $p_l=L(\rho) = \sum_a \lambda_a p_a$, and the projection onto the the CLS is
\begin{equation}
\2P_C(\rho) = \sum_a \lambda_a \sqrt{p_a(1-p_a)}
	\Big(\ketbra{\psi_{1,a}}{\psi_{2,a}}
	+\ketbra{\psi_{2,a}}{\psi_{1,a}}
	\Big). \nonumber
\end{equation}
Using the triangle inequality we have that the trace norm of $\2P_C(\rho)$ is upper-bounded by
\begin{align}
\| \2P_C(\rho)\|_1 
	&\le \sum_a \lambda_a \sqrt{p_a(1-p_a)} \left\|\ketbra{\psi_{1,a}}{\psi_{2,a}}
	+\ketbra{\psi_{2,a}}{\psi_{1,a}}
	\right\|_1 \nonumber\\
	&\le \sum_a 2\lambda_a \sqrt{p_a(1-p_a)}. \nonumber
\end{align}
Now, by the concavity of $f(x)=\sqrt{x}$ we have
\begin{align*}
\sum_a 2\lambda_a \sqrt{p_a(1-p_a)} 
	&\le 2\sqrt{\sum_a \lambda_a \,p_a(1-p_a)} \\
	&= 2\sqrt{p_l - \sum_a \lambda_a p_a^2},
\end{align*}
and by convexity of $g(x)=x^2$ we have
\begin{equation}
\sum_a \lambda_a p_a^2 \ge \left(\sum_a \lambda_a p_a\right)^2 = p_l^2,
\end{equation}
hence 
\begin{equation}
2\sqrt{p_l - \sum_a \lambda_a p_a^2} \le 2\sqrt{p_l(1-p_l)}
\end{equation}
and so $\|\2P_C(\rho)\|_1$ is upper-bounded by $2\sqrt{p_l(1-p_l)}$. \qed

\section{Proof of \cref{prop:coh-leakage-bound}}\label{proof:coh-leakage-bound}

To bound the coherent leakage rate we start with the coherence of leakage bound from \cref{prop:coherence-bound} for the output state $\2E(\ketbra{\psi_1}{\psi_1})$: 
\begin{align}
CL_1(\2E) 
	&= \int d\psi_1 \| \2P_C\2E(\ketbra{\psi_1}{\psi_1}) \|_1 \\
	&\le 2\int d\psi_1 \sqrt{p_l(\psi_1)-p_l(\psi_1)^2}
\end{align}
where $p_l(\psi_1) \equiv L(\2E(\ketbra{\psi_1}{\psi_1}))$. Next, by the concavity of $\sqrt{x}$ we have
\begin{align}
CL_1(\2E) &\le 2\sqrt{\int d\psi_1 p_l(\psi_1)-\int d\psi_1 p_l(\psi_1)^2}\\
 &= 2\sqrt{L_1(\2E)-\int d\psi_1 p_l(\psi_1)^2}.
\end{align}
For the remaining term we can rewrite it as
\begin{align}
p_l(\psi_1)^2 
	&=\Tr[\id_1\2E(\ketbra{\psi_1}{\psi_1}]^2\\
	&=\Tr[\id_2\otimes\id_2 (\2E\otimes\2E)(\ketbra{\psi_1}{\psi_1}^{\otimes 2})].
	\label{eq:pl-sq}
\end{align}
Using the result that the average over $\ketbra{\psi_1}{\psi_1}^{\otimes n}$ is given by
\begin{equation}
\int d\psi_1 \ketbra{\psi_1}{\psi_1}^{n} = \frac{\Pi_\text{sym}}{\Tr[\Pi_\text{\=sym}]}
\end{equation}
where $\Pi_\text{sym}$ is the projector on the the symmetric subspace of $\2X_1^{\otimes n}$, we may then evaluate for the case $n=2$ to obtain
\begin{equation}
\int d\psi_1 \ketbra{\psi_1}{\psi_1}^{\otimes 2}
 = \frac{\id_1\otimes\id_1 + U_{\text{SWAP}_1}}{d_1(d_1+1)}.\label{eq:psi-sq-int}
\end{equation}
Let $\{A_j/\sqrt{d_1}\}$, with $A_0 = \id_1$ be an orthonormal operator basis for $\2L(\2X_1)$. Then we may rewrite the SWAP unitary as
\begin{equation}
U_{\text{SWAP}_1} = \frac{\id_1\otimes\id_1}{d_1} + \sum_{j=1}^{d_1-1} \frac{A_j\otimes A_j}{d_1}
\end{equation}
and so \cref{eq:psi-sq-int} becomes
\begin{equation}
\int d\psi_1 \ketbra{\psi_1}{\psi_1}^{\otimes 2}
 = \frac{\id_1}{d_1}\otimes\frac{\id_1}{d_1} 
 	+\sum_{j=1}^{d_1-1} \frac{A_j\otimes A_j}{d_1^2(d_1+1)}.
	\label{eq:psi-sq-int-2}
\end{equation}
Hence by return to \cref{eq:pl-sq} we have 
\begin{align*}
\int d\psi_1 p_l(\psi_1)^2 
	&= L_1(\2E)^2
		+\sum_{j=1}^{d_1-1} 
		\frac{Tr[\id_2 \2E(A_j)]^2}{d_1^2(d_1+1)}\\
	&\ge L_1(\2E)^2
\end{align*}
Thus we obtain the result
\begin{align}
CL_1(\2E)  &= 2\sqrt{L_1(\2E)-\int d\psi_1 p_l(\psi_1)^2}\\
 &\le 2\sqrt{L_1(\2E)\big(1-L_1(\2E)\big)}.
\end{align}
The result for seepage follows the same argument. \qed

\section{Proof of \cref{prop:dle-leak}}\label{proof:dle-leak}

Let $\2E_1\in C(\2X_1)$ be a CPTP map, and $\2E_L$ be DLE of $\2E_1$. The state leakage of an initial state $\rho$ after $m$ applications of $\2E_L$ is given by
\begin{equation}
L(\2E_L^m(\rho)] = \Tr[\id_2 \2E_L^m(\rho)] = \Tr[\rho(\2E_m^\dagger)^m(\id_2)]
\end{equation}
where the adjoint channel $\2E_L^\dagger$ is given by
\begin{equation}
\2E_L^\dagger = (1-L_1) \2E_1^\dagger + L_1 \2D_{12} + L_2 \2D_{21} +(1-L_2)\2D_2
\end{equation}
Since $\2E_1$ is TP, the adjoint-channel $\2E_1^\dagger$ is unital on the computational subspace ($\2E_1^\dagger(\id_1)=\id_1$)~\cite{Wood2015qic}. Hence

Since the initial input is $\id_2$
\begin{align*}
\2E_L^\dagger\left(\alpha \id_1 + \beta \id_2\right) 
	=& \big[(1-L_1)\alpha + L_1\beta\big]\id_1 
	\\&+ L_1 \beta + (1-L_2)\beta\big]\id_2
\end{align*}
we can represent the superoperator for $\2E_L^\dagger$ with respect to the basis $\dket{\id_1},\dket{\id_2}$ as a $2\times 2$ matrix
\begin{align}
\2S_{\2E^\dagger_L} &=\begin{pmatrix} 1-L_1 & L_1 \\ L_2 & 1-L_2\end{pmatrix}.
\end{align}
Hence we can compute the $m^{\text{th}}$ power of $\2S_{\2E_L}$ obtaining
\begin{align*}
\2S_{\2E_L^\dagger}^m 
	=& \frac{1}{L_1+L_2}
	\begin{pmatrix} L_2 & L_1 \\ L_2 & L_1 \end{pmatrix}
	\\&
	+\frac{1}{L_1+L_2}\begin{pmatrix} L_1 & -L_1 \\ -L_2 & L_2 \end{pmatrix}(1-L_1-L_2)^m,
\end{align*}
and hence
\begin{align*}
\2S_{\2E_L^\dagger}^m (\id_2)
=&\left(\frac{L_1}{L_1+L_2}\right)\id
  - \frac{(1-L_1-L_2)^m}{L_1+L_2}(L_1\id_1 - L_2\id_2).
\end{align*}
Thus we have that 
\begin{align*}
L(\2E_L^m(\rho)) 
=&\left(\frac{L_1}{L_1+L_2}\right) \\&
  - \left(\frac{L_1\Tr[\id_1\rho] - L_2\Tr[\id_2\rho]}{L_1+L_2}\right)(1-L_1-L_2)^m \\
 =&\frac{L_1}{L_1+L_2} -\left(\frac{L_1}{L_1+L_2}-p_l\right)(1-L_1-L_2)^m
\end{align*}
where $p_l = L(\rho)$. \qed

\section{Proof of \cref{prop:lindblad-leakage}}\label{proof:lindblad-leakage}

To prove the result of the second order expansion of $\2S = e^{\Delta t(\2H+\2D)}$ we must show that the term
\begin{equation}
\dbra{\id_2}(\2D\2H + \2H\2D)\dket{\id_1}= \dbra{\id_1}(\2D\2H + \2H\2D)\dket{\id_2}=0.
\end{equation}
Now $\dbra{\id_i}\2H\2D\dket{\id_j} = \sum_k \dbra{\id_i}\2H \2D[A_k]\dket{\id_j}$
where we restrict ourselves to $k$-photon ladder operators of the form
\begin{equation}
A_k = \sum_{s} \alpha_s \ketbra{s\pm k}{s}.
\end{equation}
Since $A_k^\dagger A_k$ is diagonal we have that $\id_i A_k^\dagger A_k \id_j =0$. Furthermore, we have that 
\begin{equation}
A_k \id_i A_k^\dagger \id_j = |\alpha_s|^2 \ketbra{s}{s}
\end{equation}
for some $s$, and similarly for $A_k^\dagger \id_i A_k\id_j$. Using this and the property that $H$ is Hermitian, we have
\begin{align}
\Tr[A_k\id_jA_k^\dagger\id_i H] = |\alpha_s|^2 \bra{s}H\ket{s} \in \bb R.
\end{align}
Finally, expanding out the original expression we have
\begin{align*}
\dbra{\id_i}\2H \2D[A]\dket{\id_j}
=& i\Tr[A\id_jA^\dagger\id_i H] 
 	-i\Tr[H\id_i A\id_jA^\dagger] \nonumber\\
	=& -2\text{Im}\Tr[A\id_jA^\dagger\id_i H] \\
	=& 0 \\
\dbra{\id_i}\2D[A]\2H\dket{\id_j}
=&-\dbra{\id_j}\2H \2D[A]^\dagger\dket{\id_i} \nonumber\\
=& -i\Tr[A^\dagger\id_iA\id_j H] 
	+i\Tr[H\id_j A^\dagger\id_iA] \nonumber\\
=& 2\text{Im}\Tr[A^\dagger\id_iA\id_j H]\\
=& 0 \label{eq:dh-term}. \qed
\end{align*}

\end{appendix}
\end{document}